\documentclass[12pt]{ociamthesis}  

\usepackage{amssymb}
\usepackage{amsmath}
\usepackage{braket}
\usepackage{setspace}
\usepackage[utf8]{inputenc}
\usepackage[english]{babel}
\usepackage{hyperref}
\usepackage{comment}
\usepackage{tensor}
\usepackage{tikz}
\usepackage{color}
\usepackage{enumerate}
\usepackage{bbm}
\usepackage{mathtools}
\usepackage{pgfplots}

\newtheorem{theorem}{Theorem}[section]

\title{An review on \\[1ex]     
        Loop Quantum Gravity}   

\author{Pablo Antonio Moreno Casares\\[1ex]}
\college{Kellogg College}

\degree{MSc in Mathematical and Theoretical Physics}     
\degreedate{Trinity 2018}         

\pgfplotsset{compat=1.14}
\begin{document}

\baselineskip=18pt plus1pt

\setcounter{secnumdepth}{3}
\setcounter{tocdepth}{3}

\maketitle                  
\begin{romanpages}          
\mbox{}
\vspace{5cm}
\begin{flushright}
For my parents, my grandparents and Pili.
\end{flushright}        
\mbox{}
\vspace{8cm}
\begin{center}
My own view is that ultimately physical laws should find their most natural expressions in terms of essentially combinatorial principles ... Thus in accordance with such a view, [there]
should emerge some form of discrete or combinatorial space-time.

-Roger Penrose, (On the Nature of Quantum Geometry).
\end{center}
\chapter*{Acknowledgments}
I would like to thanks in first place my tutor, Prof. Lionel Mason, for his readiness in helping me with any doubts I have had during the dissertation, and for offering himself to tutor it before even the course started. He is a great tutor.

I would also like to thank all my friends in Oxford, in special Irina, Mateo, Siyu, Ziyan, Suvajit and Hemani for their supper interesting discussions on Physics and for making the course much more enjoyable. It is a real pleasure to be with such smart guys. I have ended up learning much more than I would have alone. Additionally, I would like to acknowledge all other class mates, lecturers and TAs for the master. It has been an amazing experience.

I feel in debt with my lecturers and former lecturers Isidro, Fernando and Luis, from the University of Extremadura, and Prof. Lionel Mason from the University of Oxford, for their recommendation letters, too.

From the LQG community I would like to thank Guillermo Mena, Carlo Rovelli and Alejandro Perez, for their interesting insights and tips for this dissertation, and to Edward Wilson-Ewing for offering me a place in his research group. I really appreciate how welcoming you have been with me, and I really hope that LQG is a valid physical theory, and your efforts are recognised. It would be super exciting.

And finally, I would like to deeply acknowledge my family and friends of Spain, for their unconditional support with the master, and to Pili, the best girlfriend I could imagine.

Many, many thanks to all of you.   
\chapter*{Abstract}\label{ch:abstract}
The aim of this dissertation is to review `Loop Quantum Gravity', explaining the main structure of the theory and indicating its main open issues. We will develop the two main lines of research for the theory: the canonical quantization (first two chapters) and spin foams (third). The final chapter will be devoted to studying some of the problems of the theory and what things remain to be developed. In chapter 3 we will also include an example of a simple calculation done in the frame of LQG: Schwarzschild black hole entropy.

\section*{Objectives}

The objectives of this dissertation are the following:

\begin{enumerate}
\item Use all courses I have taken during the MSc, like those related to GR or to QFT, to understand one particular approach to quantum gravity.

\item Review the main structure of Loop Quantum Gravity research project, in its two main lines: the canonical and the covariant formulation.

\item Understand some problems of the theory and what remains to be done to check if it is a real physical theory.

\item Calculate something using the LQG framework. In this case I will focus on the entropy of a Schwarzschild black hole.
\end{enumerate}


\tableofcontents            
\section*{List of symbols}

\begin{center}
\begin{tabular}{ p{3cm}  p{12cm} }
 $a,b,c...$& Spatial indices.\\
 $i,j,k...$ & Internal $su(2)$ indices.\\
 $\alpha$, $\Gamma$ & Graph.\\
 $A^i_a$ & Connection 1-form. \\ 
 $\mathcal{A}$ & Space of connections. \\  
 $\overline{\mathcal{A}}$ & Closure of the space of connections, including distributions.\\
 $A_S$ & Classical area operator.\\
 $\hat{A}_S$ & Quantum area operator.\\
 $\mathcal{C}(N), H(N)$ & Hamiltonian constraint.\\
 $\vec{\mathcal{C}}(\vec{N}), \vec{D}(\vec{N})$ & Diffeomorphism constraint.\\
 $\vec{\mathcal{C}}_G(\vec{\Lambda}), \vec{G}$& Gauss constraint.\\
 $Cyl$ & Space of cylindrical functions on $\mathcal{A}$.\\
 $Cyl^\star$ & Linear functionals of $Cyl$.\\
 Diff(M) & Space of diffeomorphisms in M.\\
 $e^a_i$ & Tetrad giving a frame of reference for each point in space $\Sigma$.\\
 $E^a_i$ & Three vector density, defined as $E^a_i:=\sqrt{\det(q)} e^a_i$.\\
 $\epsilon_{ijk}, \epsilon^{ijk}$ & Levi-Civita symbol.\\
 $\eta$ & Diffeomorphism group averaging map.\\
 $F^i_{ab}$ & Curvature of $A^i_a$ defined in \eqref{F curvature}.\\
 $G$ & Newton constant.\\
 $h_e[A]$ & Holonomy along $e$ for connection $A$. Sometimes written as $A(e)$.\\
 $\gamma$ & Immirzi parameter.\\
 $\mathcal{H}$ & Hilbert space of cylindrical functions.\\
 $\hat{J}^{(v,e)}_j$ & Angular momentum operator defined in \eqref{angular momentum operator}.\\
 $k$ & $k=8\pi G$.\\
 $\kappa(S,e)$ & $0,\pm 1 $ depending on the orientation of edge $e$ and surface $S$.\\
 $K_{ab}, K_a^i$ & Extrinsic curvature and densitized curvature, defined on \eqref{K definitions}.\\
 $\mathbf{M}$ & Master constraint.\\
 $N, \vec{N}$ & Lapse and shift respectively.\\
 $P^a_i$ & Momentum field canonical conjugate to $A^i_a$, defined in \eqref{normalized E}.\\
 $P(S,f)$ & Flux $P^a_i f^i$ across surface $S$.\\ 
  $\hat{P}(S,f)$ & Quantum flux operator across surface $S$.\\ 
 $q_{ab}$ & Metric tensor on $\Sigma$.\\
 $\Sigma$ & Slice of space. We split the manifold as $M=\Sigma \times \mathbb{R}$.\\
 $tr$ & Trace.\\
 $V$ & Classical volume defined by $q_{ab}$.\\
 $\hat{V}$ & Quantum volume operator defined by $q_{ab}$.\\
 $\hat{Y}^{(v,e)}_j$ & Spin operator defined on \eqref{Y operator}.\\
 $Y_{\gamma}$ & Function that carries unitary representations of $SU(2)$ to $SL(2,\mathbb{C})$. Defined on \eqref{Y map}.

\end{tabular}

\end{center}

\end{romanpages}            
\doublespacing
\chapter{Introduction}\label{ch:Introduction}
The search for a quantum theory of gravity is a challenge that physicists started a long time ago. Soon after Heisenberg discovered his uncertainty relations, Landau published a paper \cite{landau1931erweiterung} where he explored how the quantum theory would apply to the electromagnetic field. He suggested that the Heisenberg relations might prevent a single component of the electromagnetic field to be measured with arbitrary precision. However, Bohr and Rosenfeld realised that he was wrong \cite{bohr1933det}. 

But a friend of Landau, Matvei Bronstein, repeated the development of Bohr and Rosenfeld in the gravitational general relativistic field, discovering that in this case uncertainty relations do in fact prevent us from measuring with arbitrary precision the field \cite{Bronstein1936a,bronstein1936quantentheorie}. This is sometimes considered the birth of research on quantum gravity. 

In the 60s John Wheeler and Bryce DeWitt made the first serious attempt to combine general the Wheeler-DeWitt equation from a canonical point of view which gave rise to the \textit{geometrodynamics} program, where they would use the spatial 3-metric as a basic variable \cite{dewitt1967quantum,wheeler1969superspace}. Unfortunately, this program encountered several major difficulties and the approach remained only formal.

But why do we care about such a theory if General Relativity and Quantum Field Theory work well alone, without speaking to each other? Because there are regimes where GR or QFT are not enough alone, and they are expected to be relevant in nature, like for instance in black holes and cosmological events at very early times.
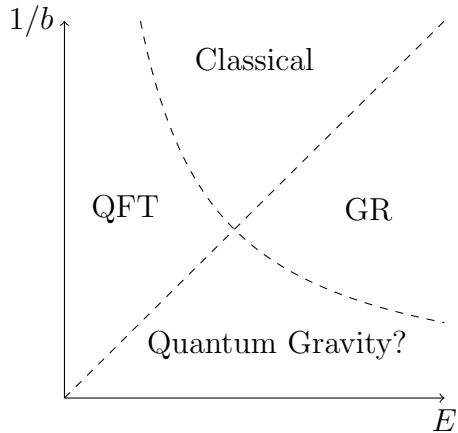
\begin{figure}
\centering
\begin{tikzpicture}[xscale=1,yscale=1,domain=0.5:5,samples=400]

    \draw[->] (0,0) -- (5,0) node[below] {$E$};
    \draw[->] (0,0) -- (0,5) node[left] {$1/b$};
    \draw[black,dashed,domain=0.0:5] plot (\x,{\x});
    \draw[black,dashed,domain=1:5] plot (\x,{5/\x});
    \node[align=left] at (0.8,2.5) {QFT};
    \node[align=left] at (2.5,4.5) {Classical};
    \node[align=left] at (4,2.5) {GR};
    \node[align=left] at (2.8,0.7) {Quantum Gravity?};
\end{tikzpicture}
\caption{Our knowledge of physical regimes, for E the energy and b the impact parameter (how close particles come to each other).}
\end{figure}

The present dissertation is a brief review of the state of Loop Quantum Gravity, the modern version of Wheeler-DeWitt old geometrodynamics. It has two main branches, the canonical, to which we will devote the first two chapters, and the covariant (the third). The canonical approach starts from the formulation of general relativity in terms of constraint algebra, and attempts quantizing it using Dirac's procedure in terms of the Ashtekar-Barbero variables. The first chapter will introduce the algebra, representation and other kinematical aspects, and the second will deal with the quantization of the constraints.

In the third chapter we will review the basics of the covariant formalism, which presents the first theory (the EPLR model) with chances of becoming the quantum theory of gravity. We will also derive the formula for the entropy of a black hole.

The last chapter will be devoted to reviewing some of the open problems of the theory. Special attention will be paid to issues raised in \cite{nicolai2005loop,nicolai2007loop}. I expect the reader to enjoy the dissertation, and to feel that although possibly completely wrong, this is a serious and beautiful attempt, and is worth studying even if only because candidates for theories of quantum gravity are scarce and much needed.
\chapter{Quantum Kinematics}\label{ch:kinematics}

\section{Classical General Relativity, without time}

Before starting the description of how to formulate quantum general relativity, let us first summarize a somewhat unusual description of general relativity, to indicate what is the general procedure we intend to carry out. Our aim is to express general relativity in a background independent way, in terms of evolution given by constraints. 

In order to have a well posed initial value problem in a manifold $M$ with a metric $g$, we need $(M,g)$ to be globally hyperbolic \cite{geroch1970}. 
This means that we can introduce a foliation of the manifold $M=\Sigma\times \mathbb{R}$. Let us decompose the Hilbert-Einstein action
\begin{equation}
S=\int_M d^{n+1}X \mathcal{L}(\Phi,\partial \Phi)= \int_M d^{n+1}x \sqrt{-g} R,\quad n=3.
\end{equation}
The phase space generated by the action will have conjugate variables $(\phi(x),\pi(x))_{x\in \Sigma}$, with $\dot{\phi}$ being de derivative with respect to the parameter $t\in \mathbb{R} $ of $M=\Sigma\times \mathbb{R}$. We now perform a Legendre transformation of the action so that we obtain
\begin{equation}
S=\int_M \mathcal{L}=\int_M \dot{\phi}\pi -\mathcal{H}=\int_M \dot{\phi}\pi -(\underbrace{ N^a D_a + NH}_{Hamiltonian}), \quad a=1,...,n,
\end{equation}
where $N^a$ and $N$ are Lagrange multipliers called shift and lapse respectively. $D_a$ and $H$ are spatial and time diffeomorphisms, and will vanish when the action is extremised, an example of Noether theorem since the action is invariant under them\footnote{Compare with $S_{QED}=\int-(1/4) F_{\mu\nu} F^{\mu\nu}- J_\mu A^\mu$ for $J_\mu$ the conserved current.}.
We define \textit{first class constraints} as those whose Poisson bracket is a linear combination of them, like our diffeomorphisms $D_a$ and $H$ \cite{HOJMAN197688}, that reproduce the Dirac algebra $\mathfrak{D}$ \cite{Thiemann:2007zz}
\begin{subequations}
\begin{align}
&\{D(\vec{N}), D(\vec{N}')\}=8\pi G D(\mathcal{L}_{\vec{N}}\vec{N}'), \label{D commutator}\\
&\{D(\vec{N}),H(N')\}=8\pi G H(\mathcal{L}_{\vec{N}}N'),\label{H D commutator}\\
&\{H(N),H(N')\}=8\pi G D(q^{-1}(NdN'-N'dN)),
\label{H commutator}
\end{align}
\label{Dirac algebra}
\end{subequations}
where $D(\vec{N})$ and $H(N)$ (diffeomorphism and Hamiltonian constraint) are the smeared versions of $D_a$ and $H$: $D(\vec{N})=\int_{\Sigma}d^3x N^a D_a$ and $H(N)=\int_{\Sigma}d^3x NH$. Finally, $\mathcal{L}_{\vec{N}}$ indicates the Lie derivative, $q$ is the determinant of the $\Sigma$-metric $q_{ab}$, and $G$ denotes Newton's constant.
The important thing to notice in \eqref{Dirac algebra} is that it is not a true Lie algebra since, due to \eqref{H commutator}, there will be phase dependence through $q$ \cite{Thiemann:2006cf}. 

We now have a phase space $\mathcal{M}$ and a collection of constraints $\{C_I\}_{I\in\mathcal{I}}=\{D(\vec{N}),H\}$, and we want to find the hypersurface $\overline{\mathcal{M}}\in\mathcal{M}$ that they annihilate. We can also form equivalence classes (orbits) $[m]:=\{m\in\mathcal{M}: m'\in [m] \iff m-m' \in  \overline{\mathcal{M}}\}$, the physically distinguishable points. The space of these orbits is called \textit{reduced phase space}. 
The idea is to find functions on $\mathcal{M}$ invariant under these gauge transformations, the \textit{Dirac observables}. We start by finding functions $T_I$ on $\mathcal{M}$ such that  $A_{IJ}:=\{C_I,T_J\}$ is (locally) invertible. Consider a new set of equivalent constraints $C_I'=\sum_{J}(A^{-1})_{IJ}C_{J}$, with $X_I$ the associated Hamiltonian vector fields, which commute weakly, i.e. on $\overline{\mathcal{M}}$ \cite{Thiemann:2006cf}. For a smooth function $f$ and real numbers $\tau_I$, respectively in the range of $T_I$, define
\begin{equation}
O_f(\tau):=[\alpha_t(f)]_{t=\tau-T}, \qquad \alpha_t(f):=\left[\exp\left(\sum_{I}t_IX_I\right)\cdot f\right].
\label{Observables}
\end{equation}

Note that $O_f(\tau)$ are weak Dirac observables \cite{Thiemann:2006cf}, which means that $\{C_I,O_f(\tau)\}|_{\overline{\mathcal{M}}}=0$, and in turn $O_f(\tau)$ are class functions, they have the same value in all the orbit. The conclusion of this ansatz is that `in general covariant systems there is no Hamiltonian, there are only Hamiltonian constraints' \cite{Thiemann:2006cf}, or equivalently, that the Hamiltonian of the system is a linear combination of constraints. 

The previous developments could give the impression that the picture of the covariant system is frozen in time and there is no evolution. But we are only talking about gauge evolution, so it is natural that (weak) Dirac observables do not change in $\tau$.
If we want to recover evolution with respect to physical time, we first see that $\alpha_t$ is a canonical transformation. Then, if we can form pairs of canonical conjugate variables $(q_a,p^a)$ and $(T_I,\pi^I)$ of the phase space, and $f$ depends only on $(q_a,p^a)$, then we can find a Hamiltonian generator for the gauge evolution in $\tau_I$ \cite{Thiemann:2004wk}. This is equivalent to finding Dirac observables $H_I(\tau)$ generating the equations of motion
\begin{equation}
\frac{\partial O_f(\tau)}{\partial \tau_I}=\{O_f(\tau),H_I(\tau(s))\}.
\label{EOM}
\end{equation}
Finally, we find a one-parameter family of diffeomorphisms $s \rightarrow \tau_I(s)$ such that 
\begin{equation}
H(s)=\sum_I\frac{d\tau_I(s)}{ds}H_I(\tau(s))
\label{Physical Hamiltonian}
\end{equation}
is independent of $s$ and positive, which is the usual Hamiltonian.

The reader is encouraged to take a look at section 2.3 in \cite{Rovelli:2014ssa}, where a nice example of this analysis  with $L=\frac{1}{2}m\dot{q}^2-V(q)$ can be found. The conclusions are that, given the Dirac algebra \eqref{Dirac algebra}, we get the physical invariants \eqref{Observables}, equations of motion \eqref{EOM}, and physical Hamiltonian \eqref{Physical Hamiltonian}. And we have done that respecting diffeomorphism invariance (we have considered all possible splittings of $M$) and without making use of the metric: in a background independent fashion.

\section{Quantization programme}
The previous discussion was possible for any general covariant theory. We would like now to focus on GR and quantize it. A priori there are two ways of doing that. One would be solving first the classical constraints, and afterwards quantizing the reduced phase space, by finding a representation of the algebra of the observables which describes their dynamics. This procedure is usually called reduced quantization, but applying it to general relativity is complicated since the algebra of the constraints is quite difficult and therefore the we cannot use the usual Fock representations.

The second way is Dirac quantization procedure \cite{Diracquant1960} and consists on quantizing the whole kinematical Hilbert space $\mathcal{H}_{kin}$ (including gauge redundancy) and promoting the classical constraints to operators $\{\hat{C}_I\}$. After that, one must find the states $\psi$ that are annihilated by the constraints, and which form the physical space $\mathcal{H}_{phys}$. This is precisely what Wheeler and DeWitt \cite{dewitt1967quantum,wheeler1969superspace} tried to carry out at least formally using the  ADM formalism \cite{ADM} .
The (formal) quantization steps that Dirac quantization prescribes to quantize general relativity are:
\begin{enumerate}[I]
\item Algebra of elementary functions $\mathfrak{E}\subset C^\infty(\mathcal{M})$.

We have to find an algebra of functions closed under the Poisson bracket and complex conjugation, where for each pair of points in $\mathcal{M}$ there exists $e\in \mathfrak{E}$ able to differentiate them (so that we are able to set up a coordinate system).
\item Quantum $^*$-algebra.

We define the $^*$-algebra $\mathfrak{V}:=\mathfrak{F}/\mathfrak{I}$, where $\mathfrak{F}=\{\lambda_1\omega_1+...+\lambda_n\omega_n, \lambda_i\in \mathbb{C}, \omega_i=(e_{i_1}...e_{i_j})\}$, and $\mathfrak{I}$ is the ideal generated by elements of the form $ee'-e'e-i\hbar\{e,e'\}$ and $e^*-\overline{e}$. $\overline{e}$ is the complex conjugate and $^*:\mathfrak{E}\rightarrow\mathfrak{E}$ is an involution, such that for any $e,e'\in \mathfrak{E}$, we have $(e+e')^*=(e^*+e'^*), \text{ }(ee')^*=e'^*e^*, \text{ }1^*=1\text{ and }(e^*)^*=e$. If $e^*=e$, it is called self-adjoint. This will be the Holonomy-Flux algebra.

\item Kinematical Hilbert space.

We will study the possible representations of the quantum algebra and although it is not the case that they will be unitarily equivalent, we shall see that under certain assumptions the Ashtekar-Lewandowski representation is unique.

\item Physical Hilbert space.

We would like to solve the constraints and find a physical inner product. This can be done by solving each of the individual constraints that we have or one single constraint called the master constraint (more on this later). 

\item Semiclassical states

Finally, we would like to find semiclassical and coherent states that reproduce GR in the appropriate limit. We will not be able to discuss this problem here, but a recent in depth discussion can be found in chapter 5 of \cite{Pullin:2017}.
\end{enumerate}

\section{Barbero-Ashtekar variables}

Let us start then by defining our phase space. We first introduce the triad which sets up a frame and a co-frame in each point, and which relate to the 3-metric of $\Sigma$ as
\begin{equation}
q^{ab}=e^a_je^b_k\delta^{jk};\quad a,b,j,k\in \{1,2,3\} \Rightarrow \sqrt{\det (q)}=\det (e):=\frac{1}{2}\epsilon^{abc}\epsilon_{ijk}e^{i}_a e^j_b e^k_c,
\end{equation}
where $i,j,k$ denote internal $so(3)=su(2)$ indices, and where it is easy to check that
\begin{equation}
e^i_a e^b_j =\delta^a_b \delta^i_j.
\end{equation}
With the triad, we can define the basic variable we will be using, the densitized triad
\begin{equation}
E^a_i:=\sqrt{det(q)}e^a_i=\frac{1}{2}\epsilon^{abc}\epsilon_{ijk}e^{j}_be^k_c \Rightarrow \det{(E)}=\sqrt{\det{(q)}}\det{(e)}=\det{(q)}
\label{densitized triad}
\end{equation}
so it is clear that 
\begin{eqnarray}
\det{(q)}q^{ab}=E^a_i E^b_j \delta^{ij}.
\end{eqnarray}
Finally we can check by substitution of \eqref{densitized triad} that 
\begin{equation}
e^i_a=\frac{1}{2}\frac{\epsilon_{abc}\epsilon^{ijk}E^b_j E^c_k}{\sqrt{|\det (E)|}} \quad \text{   and   } \quad e^a_i=\frac{sgn(\det(E)) E^a_i}{\sqrt{|\det (E)|}}
\end{equation}
hold. We also need to define the conjugate of the densitized triad $K^i_a$ making use of the extrinsic curvature $K_{ab}$ ($n$ being the unit normal to $\Sigma$)
\begin{equation}
K^i_a:=\frac{1}{\sqrt{det(E)}}K_{ab}E^b_j\delta^{ij}, \quad K_{ab}:=\frac{1}{2}\mathcal{L}_n q_{ab}.
\label{K definitions}
\end{equation}
There is an associated natural $so(3)$-connection called spin connection $\Gamma^i_a$, defined by the Cartan structure equation \cite{Perez:2004hj}
\begin{equation}
\partial_{[a}e^i_{b]}+\epsilon^i_{jk}\Gamma^j_{[a}e^k_{b]}=0 \Rightarrow \Gamma^i_a=\frac{-1}{2}\epsilon^{ij}_ke^b_j(\partial_{[a}e^k_{b]}+\delta^{kl}\delta_{ms}e^c_le^m_a\partial_be^s_c)
\label{Spin connection}
\end{equation}
Finally, we define a new connection variable $A^i_a$ as
\begin{equation}
A^i_a:=\Gamma^i_a-\gamma K^i_a,
\label{connection}
\end{equation}
where $\gamma$ is called the \textit{Immirzi parameter}.  The Poisson brackets of $E^i_a$ and $A^i_a$ are \cite{Perez:2004hj}
\begin{equation}
\{E^a_i(x),A^j_b(y)\}=k\gamma\delta^j_i\delta^a_b\delta(x-y)\quad \text{ and } \quad \{E^a_i(x),E^b_j(y)\}=0=\{A^i_a(x),A^j_b(y)\},
\end{equation}
for $k=8\pi G$. If we want to normalize the expression we just substitute $E^a_i$ with
\begin{equation}
P^a_i:=\frac{E^a_i}{k\gamma}.
\label{normalized E}
\end{equation}
Therefore $\{P^a_i,A^i_a\}$ are canonical conjugate variables. Note though, that using these variables introduces a redundancy, as the formerly 6 independent entries of the metric $q_{ab}$ now become 9 variables $E^a_i$.  We can understand this redundancy as the local symmetry $SO(3)$ that allows us to choose a preferred frame $e^i_a$. That implies that apart from the diffeomorphism and Hamiltonian constraint, we need to impose a new constraint, that will arise from \eqref{K definitions}, as it is clear that $q_{[ab]}=0\Rightarrow K_{[ab]}=0$. This means that, 
\begin{equation}
G_i=\epsilon\indices{_{ij}^k}K^j_aE^a_k=\epsilon\indices{_{ij}^k}\frac{1}{\sqrt{\det{(E)}}}K_{ab}\delta^{jl}E^b_lE^a_k=0,
\label{rotational constraint}
\end{equation}
since $\delta^{l}_jE^b_lE^a_k$ is symmetric in $j,k$ and $\epsilon\indices{_{i}^{jk}}$ antisymmetric. \eqref{rotational constraint} will be called the \textit{rotational constraint}, and in terms of $(E^a_i,A^j_b)$ we get the \textit{Gauss constraint} \cite{Pullin:2017}
\begin{equation}
G_i=\partial_aE^a_j+\epsilon\indices{_{ij}^k}A^j_aP^a_k=:\mathcal{D}_aP^a_i.
\end{equation}
The other two (Hamiltonian $\mathcal{C}$ and diffeomorphism $\mathcal{C}_a$) constraints can be obtained from the change of variables of the Hilbert-Einstein action as can be seen in \cite{Pullin:2017}:
\begin{gather}
\mathcal{C}_a =F^j_{ab}P^b_j,\label{classical diffeomorphism constraint}\\
\mathcal{C}=\frac{k\gamma^2}{2}\frac{\epsilon_j^{mn}P_m^aP_n^b}{\sqrt{det(q)}}(F^j_{ab}-(1+\gamma^2)\epsilon^{jkl}K^k_aK^m_b),
\end{gather}
with $F^j_{ab}$ is the curvature associated to $A^j_a$
\begin{equation}
F^j_{ab}:=\partial_a A^j_b - \partial_b A^j_a+\epsilon^{jkl}A^k_aA^l_b.
\label{F curvature}
\end{equation}
The action becomes \cite{Pullin:2017}
\begin{equation}
S=\int_\mathbb{R}dt\int_{M}d^3x (\dot{A}^j_aP^a_j-\underbrace{(\Lambda^jG_j+N\mathcal{C}+N^a\mathcal{C}_a)}_{\text{Hamiltonian H}}),
\end{equation}
for $\Lambda^j$ a smearing field, and $N$ and $N^a$ the lapse function and shift vector, respectively. This implies that the `Hamiltonian' is a linear combination of the constraints,
and generates the equations of motion \cite{Pullin:2017}
\begin{equation}
\dot{A}^j_a(x)=\{A^j_a,H\},\qquad \dot{P}_j^a(x)=\{P_j^a,H\}.
\end{equation}
These equations of motion, together with the vanishing condition of the constraints, are equivalent to the vacuum Einstein's equations. 
We will treat the inclusion of matter in the last chapter.
\section{Holonomy-Flux Algebra.}
Now we want to define an quantum algebra of elementary functions. However, due to the fact that general relativity is a field theory, we need to quantize smeared versions of $A^j_b,P^a_i$ instead of themselves, with a smearing that is metric-independent. In analogy to lattice quantum field theory we will make use of the holonomies
\begin{equation}
h_e[A]=\mathcal{P}\text{exp}\left(\int_e A\right)=\mathbbm{1}_{2}+\sum_{n=0}^{\infty}\int_0^1ds_1\int_{s_1}^1ds_2...\int_{s_{n}-1}^1ds_n A(e(s_1))...A(e(s_n)),\label{holonomy}
\end{equation}
where $e$ is the path on $M$ called `edge' ($e:s\in[0,1]\rightarrow e(s)  \in M$), $\mathcal{P}\text{exp}$ is the path order exponential, and we also define  $A(e(s_i)):=A^j_a(e(s_i))\tau_j e^a(s_i)$, for $\tau_j$ generator of $su(2)$. This holonomy (also denoted sometimes $A(e)$, where we are not indicating that it carries a representation of $SU(2)$ although it does) is the unique solution to 
\begin{equation}
\frac{d}{dt}h_e[A,t]=h_e[A,t]A^j_a(e(t))\tau_j\dot{e}^a(t),\quad \text{with} \quad h_e[A,0]=\mathbbm{1}_2.
\end{equation}
So, given that $A^j_a(e)$ is a $SU(2)-$connection, we can see that $h_e[A,t]\in SU(2)$ and acts as a map that parallel transports along the edge $e$ \cite{giesel2012classical}. Some of the main properties of the holonomy are \cite{Perez:2004hj} that its value is independent of the parametrization of $e$ and
\begin{equation}
h_{e_1}[A]\circ h_{e_2}[A]=h_{e_1}[A]h_{e_2}[A] \Rightarrow h_{e^{-1}}[A]= h^{-1}_{e}[A].
\end{equation}
Also, the transformation under gauge transformation $g$ is key
\begin{equation}
h_{e}[A]\rightarrow h_{e}[A]'=g(e(0))h_{e}[A]g(e(1))^{-1},
\label{gauge transformation connection}
\end{equation}
and under the action of a diffeomorphism $\phi$ (being $\phi ^*$ its pull-back)
\begin{equation}
h_{e}[\phi^* A]=h_{\phi^{-1} (e)}[A].
\end{equation}

Finally, we can also find a smearing of $P^a_j$. As in the case of the connection we were integrating along an edge, it is now natural to integrate over a surface $S$ with a $su(2)-$valued smearing field $f^j$, and we will name the result as electric flux
\begin{equation}
P(S,f)=\int_S f^j(^\star P)_j=\int_S f^j\epsilon_{abc}P^a_j dx^b\wedge dx^c,
\end{equation}
where we use the definition of the Hodge star in d dimensions:
\begin{equation}
(\alpha)_{a_1...a_p} \in \Omega^p \rightarrow  ^\star(\alpha)_{a_{p+1}...a_{p+d}}=\frac{1}{p!}\epsilon_{a_1...a_d}\alpha^{a_1...a_p} \in \Omega^{d-p}
\label{Hodge}
\end{equation}
It is now natural to analyse the Poisson bracket between $P(S,f)$ and $h_e[A]$. It will clearly depend on the relative position of $e$ and $S$, so we have the following cases
\begin{itemize}
\item If $e \cap S=e$, then we call $e$ of type `in'.
\item If $e \cap S=\emptyset$, we call $e$ of type `out'.
\item If $e$ is `above' $S$ with respect to the integration orientation, then $e$ is `up'.
\item Similarly, if $e$ is completely `under' $S$, then it is `down'.
\end{itemize}
Finally, for the case where $e$ is neither `in' or `out' we distinguish the special cases where the intersection point $p$ is the beginning of $e$, $b(e)$, or the final point $f(e)$. Then, any edge can be decomposed such that it only pierces $S$ in its initial or final points and \cite{Pullin:2017,nicolai2005loop}
\begin{equation}
\{h_e[A],P(S,f)\}=-\frac{\kappa(S,e)}{2}\times\left\{
                \begin{array}{ll}
                  +h_e[A]\tau_j f^j(b(e)) \quad\text{if} \quad e \cap S=b(e), \\
                   -\tau_j f^j(f(e))h_e[A] \quad\text{if} \quad e \cap S=f(e),
                \end{array}
              \right.
\end{equation}
where $\kappa(S,e)$ is $+1$ for type `up', $-1$ for type `down' and $0$ for types `in' and `out'.

Now, our next objective is to build a measure in the kinematical space that allows us to define an inner product, much like in quantum mechanics. However, also like in canonical quantum field theory, this requires to extend our \textit{space of smooth connections} $\mathcal{A}$, to one that also includes distributions $\mathcal{\overline{A}}$, called \textit{quantum configuration space}. We therefore want to select a classical Poisson algebra such that it can be easily extended from $\mathcal{A}$ to $\mathcal{\overline{A}}$. 

To do that we introduce in first place the notion of graph $\alpha=\{e_j, j=1,...,n; e_i\in M\}$ such that the edges only intersect in their starting or final points. The set of edges of a graph will be denoted by $E(\alpha)$ and the vertices by $V(\alpha)$. Next, we define a map
$I_E:\mathcal{A}_\alpha \rightarrow SU(2)^n$ where $ A \rightarrow (h_{e_1}[A],...,h_{e_n}[A]).$
Then let $F_\alpha$ be an arbitrary $C^\infty-$function $F_\alpha:SU(2)^n\rightarrow\mathbb{C}$. We say a function $f$ is cylindrical if it can be written as $f_\alpha(A)=F_\alpha (I_E(A))$, for some graph $\alpha$. We define
\begin{equation}
Cyl:= \bigcup_\alpha Cyl_\alpha/\sim,
\end{equation}
where $f_\alpha\sim f'_{\alpha'}$ iff $f_{\alpha''}=f'_{\alpha''}$ in all larger graphs $\alpha'': \alpha, \alpha' \subset \alpha''$.

To complete the Poisson algebra we must discuss the conjugate variables associated to the smooth cylindrical functions $Cyl$. These will be smooth vector fields $X(S,f)\in V(Cyl)$, defined by \cite{Pullin:2017}
\begin{equation}
\begin{split}
&(X(S,f)f_\alpha)(A):=\{f_\alpha,P(S,f)\}(A)\\
&=\sum_{e\in E(\alpha)}\frac{\kappa(e,S)}{2}\frac{\partial F_\alpha(h_{e}[A])}{\partial h_{e}[A]}
\times\left\{
                \begin{array}{ll}
                  +h_{e}[A]\tau_j f^j(b(e)) \quad\text{if} \quad e \cap S=b(e). \\
                   -\tau_j f^j(f(e))h_{e}[A] \quad\text{if} \quad e \cap S=f(e).
                \end{array}
              \right.
\end{split}
\end{equation}
We can conclude that the classical Poisson algebra (flux-holonomy algebra $\mathfrak{V}$) is formed by the $^*$-subalgebra $Cyl \times V(Cyl)$.
\section{The Ashtekar-Lewandowski representation.}
As we mentioned earlier on, our next task is to construct a measure on $\mathcal{\overline{A}}$. This implies adding distributions to our classical space $\mathcal{A}$. In particular we want to find a space $\mathcal{\overline{A}}$ such that $\mathcal{H}$ will be isomorphic to some $L_2-$space over $\mathcal{\overline{A}}$ with some yet to be defined measure on $\mathcal{\overline{A}}$. As a first step, for a given graph $\alpha$ we define, for not necessarily smooth connections $A(e(s_i))$, the measure 
\begin{equation}
\braket{f_\alpha, \tilde{f}_\alpha}:=\int_{SU(2)^n}\prod_{i=1}^n d\mu_{H}(h_{e_i}[A])\overline{F_\alpha(h_{e_1}[A],...,h_{e_n}[A])}F_\alpha(h_{e_1}[A],...,h_{e_n}[A]),
\end{equation}
thanks to the fact that we have a natural measure $d\mu_{H}(g)$ (the Haar measure) over $SU(2)$. The Hilbert spaces are defined as $\mathcal{H}_{\alpha}:=L_2(\overline{\mathcal{A}}_\alpha, d\mu_\alpha)$. To construct the so called Ashtekar-Lewandowski measure, what we need to do is to extend this concept to the case where they live on different graphs. 
\begin{equation}
\braket{f_{\alpha'}, \tilde{f}_{\alpha''}}:=\int_{SU(2)^n}\prod_{i=1}^n d\mu_{H}(h_{e_i}[A])\overline{F_\alpha(h_{e_1}[A],...,h_{e_n}[A])}F_\alpha(h_{e_1}[A],...,h_{e_n}[A]),
\label{scalar product}
\end{equation}
where $\alpha',\alpha''\subset \alpha$, and the Hilbert space will be $\mathcal{H}=L_2(\overline{\mathcal{A}}, d\mu_{AL})$.

Now that we have a (kinematical) Hilbert space, we want to find a representation $\pi$ of the algebra. To do that we have to define how the holonomy and flux vectors operate on $\mathcal{H}$, that can be done as $Cyl$ is dense on $\mathcal{H}$. The holonomy operator will act multiplicatively and the flux vector fields as derivation operators. For $\psi\in\mathcal{H}$,
\begin{align}
(\pi(f)\psi)(A)&:=(\hat{f}\psi)(A)=f(A)\psi(A),\label{holonomy operator}\\
(\pi(P(S,f))\psi)(A)&:=\hat{P}(S,f)(\psi(A))=(X(S,f)\psi)(A).
\end{align}
We also define left/right invariant operators. Given $f:SU(2)\rightarrow \mathbb{C}$ and $g\in SU(2)$,
\begin{equation}
(L_j f)(g):=\frac{d}{dt}(f(g e^{t\tau_j}))_{t=0}, \qquad (R_j f)(g):=\frac{d}{dt}(f(e^{t\tau_j}g))_{t=0}.
\label{right and left invariant vector fields}
\end{equation}
Finally, we the flux operator will act on $f_\alpha\in Cyl_\alpha$ as \cite{Pullin:2017}
\begin{equation}
\hat{P}(S,f)f_\alpha(A)=\frac{\hbar}{2}\sum_{v\in V(\alpha)}f^j(v)\sum_{\substack{e\in E(\alpha)\\
                  e\cap v \neq \emptyset}}
                  \kappa(e,S)\hat{Y}^{(v,e)}_j f_{\alpha}(A),
\label{flux operator on cylindrical function}
\end{equation}
with
\begin{equation}
\hat{Y}^{(v,e)}_j:=\mathbbm{1}_\mathcal{H}\times...\times \mathbbm{1}_\mathcal{H} \times \left\{
\begin{array}{ll}
+i R^e_j\\
-i L^e_j
\end{array}
\right\}
\times \mathbbm{1}_\mathcal{H}\times...\times \mathbbm{1}_\mathcal{H}, \quad \text{if} \quad 
\left\{
\begin{array}{ll}
\text{e outgoing at v.}\\
\text{e ingoing at v.}
\end{array}
\right\}
\label{Y operator}
\end{equation}

So far we have found one possible representation of the Flux-Holonomy algebra. The question now is if we can find a different representation leading to a different quantum theory. In principle, we could think that maybe some analogue of the Stone-Von Neumann theorem \cite{stone1930linear,neumann1931eindeutigkeit} holds so that all representations are unitarily equivalent. However, from \eqref{scalar product}, we will see in the next section that  $\braket{f_{\alpha'}, \tilde{f}_{\alpha''}}=0$ unless $\alpha'=\alpha''$ \cite{nicolai2005loop}. This implies that operators cannot be weakly continuous, therefore violating one of the assumptions of the Stone-Von Neumann theorem. However, not everything is lost. In particular, there exists one theorem (usually called LOST-theorem \cite{lewandowski2006uniqueness,fleischhack2009representations}) that indicates that under certain reasonable assumptions, the Ashtekar-Lewandowski representation is unique.
\begin{theorem} There is only one cyclic representation of the holonomy-flux algebra $\mathfrak{V}$, with a diffeomorphism invariant cyclic vector: the Ashtekar-Lewandowski representation.
\end{theorem}
A representation is \textit{cyclic} if all vectors in $\mathcal{H}$ are cyclic, and a \textit{cyclic vector} $\Omega$ is a vector such that $\{\pi(a)\Omega|a\in\mathfrak{V}\}$ ($\pi$ the holonomy operator \eqref{holonomy operator}) is dense in $\mathcal{H}$. It implies that the representation is irreducible \cite{Pullin:2017} and some consequences are the existence of geometric operators with discrete area, and the fact that only finite diffeomorphism have unitary operators associated. More on this can be found in \cite{Thiemann:2007zz}, chapter 8.

\section{Spin networks as an orthonormal basis of $\mathcal{H}_{kin}$}

Having defined the cylindrical functions and a scalar product we want to find an orthonormal basis of $\mathcal{H}_{kin}$. Let us denote by $\pi^j$ the finite dimensional, unitary, irreducible representations of $SU(2)$. Define, for $g\in SU(2)$
\begin{equation}
b^j_{mn}:SU(2)\rightarrow\mathbb{C}, \quad g \mapsto \braket{g|b^j_{mn}}:=\sqrt{dim(\pi^j)}\pi^j_{mn}(g),\quad m,n=1,...,dim(\pi^j).
\end{equation}
This will allow us to define an inner product, making use of Haar measure $d\mu_H$
\begin{equation}
\braket{b^j_{mn},b^{j'}_{m'n'}}:=\int_{SU(2)} d\mu_H (g)\sqrt{2j+1}\pi^j_{mn}(g)\sqrt{2j'+1}\pi^{j'}_{m'n'}(g)=\delta^{jj'}\delta^{nn'}\delta_{nn'},
\end{equation}
where the last equality is the Peter and Weyl theorem\footnote{It is the $SU(2)$ equivalent of $f(\theta)=\sum_n f_n \exp(i\theta n)$ for $f\in L^2(U(1))$, with $f_n=(2\pi)^{-1}\int d\theta f(\theta) \exp(-i \theta n)$. $\exp(i\theta n)$ are the unitary irreducible representations of $U(1)$.} (a proof can be seen in  chapter 31.2 of \cite{Thiemann:2007zz}), that says that {$b^j_{mn}$} form an orthonormal basis of $\mathcal{H}=L_{2}(SU(2),d\mu_H)$. We can use this to define an orthonormal basis of $\mathcal{H}_{kin}=L_2(SU(2)^n,d\mu_{AL})$, the spin network functions (SNF) of a certain graph $\alpha$.
\begin{equation}
\ket{s^{\vec{j}}_{\alpha,\vec{m}\vec{n}}}: \quad\overline{\mathcal{A}}_{\alpha}\rightarrow\mathbb{C}\quad
 A\mapsto \braket{A|s^{\vec{j}}_{\alpha,\vec{m}\vec{n}}}:=\prod_{i=1}^n\sqrt{2j_{e_i}+1}\pi^{j_{e_i}}_{m_{e_i}n_{e_i}}(h_{e_i}[A])
\label{SNF}
\end{equation}
Then, we can see that $\mathcal{H}_{kin}=\bigoplus_\alpha \mathcal{H}_\alpha$, and its orthonormal basis is $\ket{s^{\vec{j}}_{\alpha,\vec{m}\vec{n}}}$. So, one important conclusion that we will use afterwards is that any two spin network functions will be orthogonal unless their graphs coincide and the assigned spin is the same for each edge.

Recall that $\hat{Y}^{(v,e)}_j$  acts as $\hat{Y}^{(v,e)}_j\ket{j_em_e}_{n_e}=\sum_{\tilde{m}_e}\pm i\pi^{j_e}_{m_e\tilde{m}_e}(\tau_j)\ket{j_e\tilde{m}_e}_{n_e}$ with the $\pm$ depending on whether the edge is `incoming' or `outgoing'. Instead of using it we can rewrite this expression in terms of the usual angular momentum operator and basis by the means of a unitary map $W: \mathcal{H}^{jm}\rightarrow \mathcal{H}_{jm}$ such that $W\hat{J}^{(v,e)}_jW^{-1}=\hat{Y}^{(v,e)}_j$ and
\begin{align}
W: \mathcal{H}^{jm}\rightarrow \mathcal{H}_{jm}, \quad \ket{jm;n}\mapsto W\ket{jm;n}=\sum_{\tilde{m}_e}\pi^j_{m_e,\tilde{m}_e}(i\sigma_2)\ket{j_e\tilde{m}_e}_{n_e},\\
W^{-1}: \mathcal{H}_{jm}\rightarrow \mathcal{H}^{jm}, \quad \ket{j_e\tilde{m}_e}_{n_e}\mapsto W^{-1}\ket{j_e\tilde{m}_e}_{n_e}=\sum_{\tilde{m}_e}\pi^j_{m_e,\tilde{m}_e}(-i\sigma_2)\ket{jm;n}
\label{W-1 map}
\end{align}
\begin{figure}
\begin{center}
\includegraphics[width=0.7\textwidth]{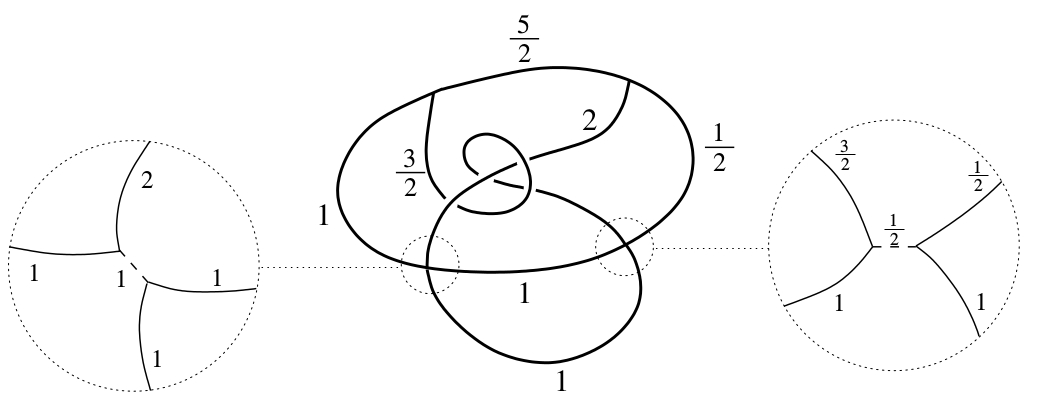}
\end{center}
\label{Spin network figure}
\caption{Graphical representation of a spin network. Any vertex with valence higher than  3 can be decomposed by adding additional edges. Figure taken from \cite{Perez:2004hj}.}
\end{figure}
From \eqref{SNF} and using \eqref{W-1 map},
\begin{equation}
W^{-1}\pi^{j_e}_{m_en_e}(h_{e}[A])=\pi^{j_e}_{m_e\tilde{m}_e}(-i\sigma_2)\frac{\braket{A|j_e\tilde{m}_e}}{\sqrt{2j_e+1}},
\end{equation}
and then substituting in \eqref{SNF} we can write 
\begin{equation}
\braket{A|s^{\vec{j}}_{\alpha,\vec{m}\vec{n}}}:=\prod_{i=1}^n\pi^{j_{e_i}}_{m_{e_i}\tilde{m}_{e_i}}(-i\sigma_2)\braket{A|j_{e_i},\tilde{m}_{e_i};n_{e_i}}.
\end{equation}
We can also write our Hilbert space in terms of abstract angular momentum numbers. The angular momentum operators are defined as
\begin{equation}
\hat{J}^{(v_i)}_j:=\sum_{\substack{e\in E(\alpha)\\ 
e\cap v_i\neq \emptyset}}\hat{J}^{(v_i,e)}_j\quad \text{and}\quad (\hat{J}^{(v_i)})^2:=\eta^{jk}\hat{J}^{(v_i)}_j\hat{J}^{(v_i)}_k,
\label{angular momentum operator}
\end{equation}
where $\eta^{jk}$ is the Killing metric for $su(2)$ and $(\hat{J}^{(v_i)})^2$ only acts non trivially on the vertex $v_i$, with eigenvalues $l_i(l_i+1)$. 
Then, in order to write $\mathcal{H}$ as a direct sum of orthogonal $\mathcal{H}_\alpha$s we need to introduce the notion of admissible graph labelling. One defines admissible labelling of a graph $\alpha$ by irreducible representations \cite{Pullin:2017} if no edge carries a trivial representation, nor does a 2-valent vertex (otherwise we would have redundancies in counting graphs). Therefore one may write.
\begin{equation}
\mathcal{H}_{kin}=\bigoplus_{\alpha}\bigoplus_{\substack{\vec{j},\vec{l}\\
                     \text{admissible}}} \mathcal{H}_{\alpha,\vec{j}\vec{l}}.
                     \label{H_kin}
\end{equation}

\section{Conclusions}
In this chapter we have seen that it is possible to rewrite General Relativity in terms of an algebra of constraints. We have started quantizing it by Dirac's procedure, and using Ashtekar variables we have rewritten our theory in terms of fluxes and holonomies. Then we found the Flux-Holonomy algebra based on $Cyl$ functions, and studied that under certain assumptions the Ashtekar-Lewandowski representation (with the corresponding inner product)  is unique. Finally, we have seen that we can find an orthonormal basis for $\mathcal{H}_{kin}$ using Spin Network Functions.
\chapter{Quantum Dynamics}\label{ch:dynamics}

In order to find the dynamics of the theory, we need to solve the `quantum Einstein equations' of LQG. For $\psi\in \mathcal{H}_{phys}$, these are (formally)
\begin{equation}
\hat{\mathcal{\vec{C}}}_G(\vec{\Lambda})\psi=0,\quad \hat{\mathcal{\vec{C}}}(\vec{N})\psi=0,\quad
\hat{\mathcal{C}}(N)\psi=0,
\end{equation}
as we saw that the physical space is given by those states that are annihilated by all three constraints. In this chapter we will review the procedure to solve these constraints and also study the geometrical kinematical operators, one of the main features of LQG.
\section{Gauss constraint}
The solution states to the Gauss constraint are those that are invariant under $SU(2)$ gauge transformations. Let us start from \eqref{SNF}, and recall how the connection transforms under gauge transformations, \eqref{gauge transformation connection}. Then
\begin{equation}
\begin{split}
\pi^j_{m_en_e}(h_{e}[A])\rightarrow &\pi^j_{m_en_e}(g(b(e))h_{e}[A]g^{-1}(f(e)))=\\
&\pi^j_{m_e\alpha_e}(g(b(e))\pi^j_{\alpha_e\beta_e}(h_{e}[A])\pi^j_{m_e\beta_e}(g^{-1}(f(e))).
\end{split}
\end{equation}
We may rewrite \eqref{SNF} as
\begin{equation}
\braket{A|s^{\vec{j}}_{\alpha,\vec{n}\vec{m}}}=\prod_{v\in V(\alpha)}\prod_{\substack{e\in E(\alpha)\\
                  e\cap v \neq \emptyset}}
                  \sqrt{2j_e+1}\pi^{j_e}_{m_en_e}(h_{e}[A]).
\end{equation}
Since the gauge transformation acts only on the vertices (see \eqref{gauge transformation connection}) let us get an invariant tensor. The Hilbert space associated to the vertex will be $\mathcal{H}_v=\bigotimes_{\substack{e\in E(\alpha)\\
                                             e\cap v\neq \emptyset}}\mathcal{H}_{j_e}$. A basis for $\mathcal{H}_v$ can be found in terms of $(n,m)-$tensors, for $n$ incoming and $m$ outgoing edges at the given vertex. Given one such tensor $\iota_{i;\beta_1...\beta_m}^{\alpha_1...\alpha_n}$, the transformation rule is
\begin{equation}
\iota_{i;\beta_1...\beta_m}^{\alpha_1...\alpha_n} \rightarrow \pi^{j_{e_1}}(g(v))\indices{^{\alpha_1}_{\gamma_1}}...\pi^{j_{e_n}}(g(v))\indices{^{\alpha_n}_{\gamma_n}}
\overline{\pi}^{j_{e_{n+1}}}(g(v))\indices{^{\delta_{n+1}}_{\beta_{n+1}}}...\overline{\pi}^{j_{e_{n+m}}}(g(v))\indices{^{\delta_{n+m}}_{\beta_{n+m}}}
\iota_{i;\delta_1...\delta_m}^{\gamma_1...\gamma_n},
\end{equation}
where $\overline{\pi}^{j_{e_{i}}}$ denotes the dual representation and $\overline{\pi}(g(v))=\pi(g^{-1}(v))^T$. We say that a tensor is invariant under gauge transformations (or \textit{intertwiner}) when, by components
\begin{equation}
\pi^{j_{e_1}}(g(v))\indices{^{\alpha_1}_{\gamma_1}}...\pi^{j_{e_n}}(g(v))\indices{^{\alpha_n}_{\gamma_n}}
\overline{\pi}^{j_{e_{n+1}}}(g(v))\indices{^{\delta_{n+1}}_{\beta_{n+1}}}...\overline{\pi}^{j_{e_{n+m}}}(g(v))\indices{^{\delta_{n+m}}_{\beta_{n+m}}}
\iota_{i;\delta_1...\delta_m}^{\gamma_1...\gamma_n}=\iota_{i;\beta_1...\beta_m}^{\alpha_1...\alpha_n}.
\end{equation}
Note that due to the identity $Hom(V,W)\simeq W\otimes V^*$, we can also see intertwiners as $\iota: \bigotimes_{k=1}^m\mathcal{H}_{j_{e_k}}\rightarrow \bigotimes_{l=1}^n\mathcal{H}_{j_{e_l}}$. Finally, given some spin network functions, we may construct invariant spin network functions by contracting the former with intertwiners to get singlets. Then, a gauge invariant spin network function for a graph with $\{e_{v_1}...e_{v_m}\}$ outgoing edges and $\{e_{v_m+1}...e_{v_m+v_n}\}$ ingoing edges at a vertex $v$ is \cite{Pullin:2017}
\begin{equation}
\begin{split}
\braket{A|s^{\vec{j}}_{\alpha,\vec{i}}}:=&\prod_{v\in V(\alpha)}\iota_v \prod_{\substack{e\in E(\alpha)\\
                  e\cap v \neq \emptyset}}
                  \sqrt{2j_e+1}\pi^{j_e}_{n_em_e}(h_{e}[A])\\
=&\prod_{v\in V(\alpha)}\iota_{v;m_1...m_{v_n}}^{n_1...n_{v_m}}\sqrt{2j_{e_1}+1} ... \sqrt{2j_{e_{v_m+v_n}}+1}\pi^{j_{e_1}}_{n_1 m_{e_1}}(h_{e_1}[A])...\\
&\pi^{j_{v_m}}_{n_{v_m}m_{e_{v_m}}}(h_{e_{v_m}}[A])\pi^{j_{e_{v_m+1}}}_{n_{e_{v_m+1}}m_1}(h_{e_{v_m+1}}[A])...\pi^{j_{e_{v_m+v_n}}}_{m_{e_{v_m+v_n}}n_{v_n}}(h_{e_{v_m+v_n}}[A]).
\end{split}
\label{gauge invariant SNF}
\end{equation}
This implies that this Hilbert space is a subset of the kinematical Hilbert space (it will not be the case for the other constraints). The subset is such that in each vertex of $\alpha$ the total angular momentum is a singlet, and then we must modify \eqref{H_kin} so that 
\begin{equation}
\mathcal{H}_{kin}^G=\bigoplus_{\alpha}\bigoplus_{\substack{\vec{j},\vec{l}\\
                     \text{admissible}}} \mathcal{H}_{\alpha,\vec{j}\vec{l}=0}.
                     \label{H_Gauss}
\end{equation}

\section{Geometrical kinematical operators}

In this section we will study several geometrical operators arising from our choice of Ashtekar variables and the flux-holonomy algebra $\mathfrak{V}$. We will start from the simplest one, the area operator, key in the derivation of the black hole entropy.

\subsection{The area operator}

The area operator has been studied in detail. For the original papers one may take a look at \cite{smolin1992recent,rovelli1995discreteness,ashtekar1997quantum}. The procedure to obtain it is first writing the classical expression in Ashtekar variables, then adding a regulator, promoting the expression to a quantum operator on $\mathcal{H}_{kin}$, and finally checking (and hoping) that when the regulator is removed, the operator is still well defined. We start from  the classical expression for the area of a surface $S$ divided into $N$ 2-cells $\{S_I\}$ in terms of the flux, and use \eqref{normalized E}	
\begin{equation}
A_S=\lim_{n\rightarrow\infty} \sum_{I=1}^N \sqrt{n_a n_b E^a_j(S_I) E^b_k (S_I)}=k\gamma\int_S d^2u\sqrt{P^\perp_j P^\perp_k \delta^{jk}}(u),
\end{equation}
where $P^\perp_j$ is the projection of $P_j^a$ onto the normal of $\Sigma$, $n_a$. We want to regulate this expression using a density function $f^\epsilon_u(u')$ that for small $\epsilon$ tends to a delta function with maximum in $u'$: $lim_{\epsilon\rightarrow 0}f^\epsilon_u(u')=\delta_u(u')$. The regulated classical expression can be defined as
\begin{equation}
[P^\perp_j]^\epsilon (u):=\int_S d^2u' f^\epsilon_u(u')P^\perp_j(u'),
\label{regularised classical momentum operator}
\end{equation}
with the property that 
\begin{equation}
\lim_{\epsilon\rightarrow 0}[P^\perp_j]^\epsilon (u)= P^\perp_j (u).
\end{equation}
The regularised area classical operator is then
\begin{equation}
[A_S]^\epsilon:=k\gamma\int_S d^2u \left|[P^\perp_j]^\epsilon (u)[P^\perp_k]^\epsilon (u)\delta^{jk}\right|^{\frac{1}{2}},
\label{regularised classical area operator}
\end{equation}
and we will promote this expression to quantum operator by substituting $P^\perp_j (u')$ in \eqref{regularised classical momentum operator} by the functional derivative $\hat{P}^\perp_j=-i\hbar\frac{\delta}{\delta A^j_\perp}$, obtaining the regularised operator $[\hat{P}^\perp_j]^\epsilon (u)$. Because of \eqref{flux operator on cylindrical function}, we can write
\begin{equation}
[\hat{P}^\perp_j]^\epsilon (u) \ket{s^{\vec{j}}_{\alpha,\vec{m}\vec{n}}}=\frac{\hbar}{2}\sum_{v\in V(\alpha)}f^\epsilon_u(v)\sum_{\substack{e\in E(\alpha)\\
                  e\cap v \neq \emptyset}}
                  \kappa(e,S)\hat{Y}^{(v,e)}_j \ket{s^{\vec{j}}_{\alpha,\vec{m}\vec{n}}},
\end{equation}
and therefore \eqref{regularised classical area operator} becomes
\begin{equation}
\begin{split}
[\hat{A}_S]^\epsilon\ket{s^{\vec{j}}_{\alpha,\vec{m}\vec{n}}}&=4\pi \gamma l_p^2 \int_S d^2u\left|\left(\sum_{v\in V(\alpha)}f^\epsilon_u(v)\sum_{\substack{e\in E(\alpha)\\
                  e\cap v \neq \emptyset}}
                  \kappa(e,S)\hat{Y}^{(v,e)}_j \right)^2\right|^\frac{1}{2}\ket{s^{\vec{j}}_{\alpha,\vec{m}\vec{n}}}\\
                  &=4\pi \gamma l_p^2 \int_S d^2u\sum_{v\in V(\alpha)}f^\epsilon_u(v)\left|\left(\sum_{\substack{e\in E(\alpha)\\
                  e\cap v \neq \emptyset}}
                  \kappa(e,S)\hat{Y}^{(v,e)}_j \right)^2\right|^\frac{1}{2}\ket{s^{\vec{j}}_{\alpha,\vec{m}\vec{n}}},
\end{split}
\end{equation}
where we have used that $f^\epsilon_u(v)f^\epsilon_{u}(v')=\delta_{v,v'}(f^\epsilon_u(v))^2$, since the functions $f^\epsilon_u(v)$ are non-zero at at most one vertex. Removing the regulator, $f^\epsilon_u(v) \rightarrow \delta_u(v)$ and
\begin{equation}
\begin{split}
\hat{A}_S\ket{s^{\vec{j}}_{\alpha,\vec{m}\vec{n}}}&:=\lim_{\epsilon\rightarrow 0}[\hat{A}_S]^\epsilon\ket{s^{\vec{j}}_{\alpha,\vec{m}\vec{n}}}\\
&=4\pi \gamma l_p^2 \sum_{\substack{v\in V(\alpha)\\
                             v\in I(S)}}\left|\left(\sum_{\substack{e\in E(\alpha)\\
                  e\cap v \neq \emptyset}}
                  \kappa(e,S)\hat{Y}^{(v,e)}_j \right)^2\right|^\frac{1}{2}\ket{s^{\vec{j}}_{\alpha,\vec{m}\vec{n}}},
\end{split}
\end{equation}
where we have only included those terms for which $\kappa(S,e)\neq 0$. That is $I(S)=\{v\in e\cap S | \kappa(e,S)\neq 0, e\in E(\alpha), v\in V(\alpha)\}$. This is the final form of the area operator. Let us nevertheless analyse it a bit further. Define
\begin{equation}
\hat{Y}^{v,u}_j:=\sum_{e\in E(v,u)} \hat{Y}^{(v,e)}_j, \qquad \hat{Y}^{v,d}_j:=\sum_{e\in E(v,d)} \hat{Y}^{(v,e)}_j,
\end{equation}
with $E(v,u)$ and $E(v,d)$ are the edges of type up and down respectively. Then,
\begin{equation}
\begin{split}
\left(\sum_{\substack{e\in E(\alpha)\\
                  e\cap v \neq \emptyset}}
                  \kappa(e,S)\hat{Y}^{(v,e)}_j \right)^2 &= \left(\hat{Y}^{v,u}_j-\hat{Y}^{v,d}_j\right)^2=(\hat{Y}^{v,u}_j)^2+(\hat{Y}^{v,d}_j)^2-2\hat{Y}^{v,u}_j\hat{Y}^{v,d}_j\\
&=2(\hat{Y}^{v,u}_j)^2+2(\hat{Y}^{v,d}_j)^2-\left(\hat{Y}^{v,u}_j+\hat{Y}^{v,d}_j\right)^2,
\end{split}
\end{equation}
where we have used that those operators commute because they do not act on the same edges. Finally, since $(R_jf)(g), (L_jf)(g)\in su(2)$  then $ \hat{Y}^{v,u}_j, \hat{Y}^{v,d}_j \in su(2)^n$, and we know that
\begin{equation}
Spec(\hat{A}_S)=4\pi\gamma l_p^2\sum_{v\in I(S)}\sqrt{2j_{u,v}(j_{u,v}+1)+2j_{d,v}(j_{d,v}+1)-j_{u+d,v}(j_{u+d,v}+1)},
\end{equation}
where we should take into account that $|j_{u,v}-j_{d,v}|\leq j_{u+d,v}\leq j_{u,v}+j_{d,v} $. So, we get the interesting result that the smallest area eigenvalue (also known as area gap) is 
\begin{equation}
 \lambda_0=2\pi \gamma l_p^2\sqrt{3}.
\end{equation}

\subsection{The volume operator}

The volume operator is another important geometrical operator of the theory, as it plays a key role in Thiemann's work on the solution of the Hamiltonian constraint. To derive it, as in the area, we start by defining the classical expression for the volume
\begin{equation}
 V_R=\int_R d^3x \sqrt{\det (q)}=(k\gamma)^{\frac{3}{2}}\int_R d^3x \sqrt{|\det (P^a_j)|}.
 \label{classical volume}
\end{equation}
The strategy is similar to that of the area operator. One starts by partitioning $\mathcal{R}$ into cubic cells $C^\epsilon$ of volume smaller than $\epsilon$ and such that there is no overlap between cells except for points in the boundary. For each cell $C^\epsilon_I$ we choose three 2-surfaces $S_I^a$ which are constant on the three spatial coordinates $x^a$. The idea is to define the volume making use of the flux through these surfaces. Choosing the usual basis of $su(2)$, $\tau^i=i\sigma^i/2$, as the smearing functions for the flux, we approximate \eqref{classical volume} as 
\begin{equation}
V_R^\epsilon=(k\gamma)^{\frac{3}{2}} \sum_{C^\epsilon_I\in\mathcal{P}^\epsilon}\sqrt{|Q_{C^\epsilon_I}|}, \quad Q_{C^\epsilon_I}:=\frac{1}{3!}\epsilon^{ijk}\epsilon^{abc}P_i(S_a)P_j(S_b)P_k(S_c),
\label{classical regulated volume operator}
\end{equation}
that can be directly promoted to an operator. The problem arises in the fact that, unlike for the area operator, when the regulator is removed, we will still have a dependence with respect to the chosen partition. More specifically, it will depend on the position of surfaces $S_I^a$ in relation to edges and vertices. For example, one may get an arbitrarily large number of cells with unit contribution, if in each cell without vertex, one edge intersects the three surfaces with $\kappa(e,S)\neq 0$, as each intersection implies multiplying the cell contribution by $\tau^i$, and $\tau^1\tau^2\tau^3=i$ \cite{nicolai2005loop}. This leads to a divergent result, so one must exclude this possibility by hand. In the same way, we must ensure that our partition allows, when $\epsilon$ sufficiently small, for the vertex $v$ to sit in the intersection between surfaces $v=S_I^a\cap S_I^b\cap S_I^c$. 

To eliminate this dependence on the relative position of edges and surfaces, one starts by substituting \eqref{flux operator on cylindrical function} on the quantum analogue of \eqref{classical regulated volume operator}
\begin{equation}
\hat{Q}_I=\frac{1}{3!}\sum_{e_i,e_j,e_k}\epsilon(e_i,e_j,e_k)\epsilon^{ijk}\epsilon^{abc}\hat{Y}^{(e_1,v)}_{i,a}\hat{Y}^{(e_2,v)}_{j,b}\hat{Y}^{(e_3,v)}_{k,c},
\label{operator Q}
\end{equation}
with 
\begin{equation}
\epsilon(e_i,e_j,e_k)=\kappa(e_i,S_I^1)\kappa(e_j,S_I^2)\kappa(e_k,S_I^3).
\end{equation}
Afterwards, we perform a group averaging 
so that instead of $\epsilon(e_i,e_j,e_k)$ we use 
\begin{equation}
\hat{\epsilon}(e_i,e_j,e_k):=\int d\mu(\theta_i,\theta_j,\theta_k) \kappa(e_1,S_I^i(\theta_i))\kappa(e_j,S_I^k(\theta_j))\kappa(e_k,S_I^k(\theta_k)),
\end{equation}
where $\theta_i$ are suitable angular coordinates, and $d\mu$ an arbitrary measure. Up to the choice of measure and therefore to a multiplicative factor $c_{reg}$, this gives a well defined answer for the volume operator. We must also impose the condition that, due to diffeomorphism invariance, $c_{reg}$ can still be arbitrary but the same for all cells.

The volume operator will act on a gauge invariant spin network function \eqref{gauge invariant SNF}, by replacing the intertwinners $\iota_v$ by \cite{nicolai2005loop}
\begin{equation}
\begin{split}
\iota_{i;\alpha_{n+1}...\alpha_{n+m}}^{\alpha_1...\alpha_n}\rightarrow &\sum_{e_ie_je_k\in \{e_1...e_{n+m}\}} \hat{\epsilon}(e_i,e_j,e_k)\epsilon^{ijk}\epsilon^{abc} \pi^{j_{e_i}}(g(v))\indices{_{\alpha_i \gamma_i}}\\ &\pi^{j_{e_j}}(g(v))\indices{_{\alpha_j \gamma_j}}\pi^{j_{e_k}}(g(v))\indices{_{\alpha_k \gamma_k}}\left(\prod_{\substack{l\in\{1..n+m\}\\
                       l\neq i,j,k}}\delta_{\alpha_l \gamma_l}\right)
\iota_{i;\gamma_{n+1}...\gamma_{n+m}}^{\gamma_1...\gamma_n}.
\end{split}
\end{equation}
We can see that the volume operator acts only at the vertices of the graph, by changing the intertwinners, but will not modify the graph.

This is one version of the volume operator, derived by Ashtekar and Lewandowski (AL) \cite{ashtekar1997volume}. However, in the literature, another expression was proposed by Rovelli and Smolin (RS) \cite{rovelli1995discreteness}
\begin{align}
&\hat{V}_{v,RS}=c_{RS}\sum_{e_i\cap e_j\cap e_k=v}\left| \hat{Q}_{ijk}\right|^{\frac{1}{2}} \label{V_RS}\\
&\hat{V}_{v,AL}=c_{AL}\left| \sum_{e_i\cap e_j\cap e_k=v} \hat{\epsilon}(e_i,e_j,e_k) \hat{Q}_{ijk} \right|^{\frac{1}{2}}\label{V_AL},
\end{align}
with $\hat{Q}_{ijk}=\epsilon^{ijk}\epsilon^{abc}\hat{Y}^{(e_i,v)}_{k,c}\hat{Y}^{(e_j,v)}_{k,c}\hat{Y}^{(e_k,v)}_{k,c}$. We can see that there is an important difference between these two operators, namely that $\hat{V}_{v,RS}$ does not take into account the orientation of the edges, what implies that it is covariant under homomorphisms. In contrast $\hat{V}_{v,AL}$ is only covariant under diffemorphisms.

The spectrum of the volume operator is much more complicated and there is no general analytical formula. However a formula for the matrix elements has been found \cite{brunnemann2006simplification}, and used to calculate volumes of vertex up to valence 7 \cite{brunnemann2008properties,brunnemann2008propertiesII,brunnemann2010oriented}. One interesting fact obtained in those works is that the presence of a volume gap depends on the factors $\hat{\epsilon}(e_i,e_j,e_k)$. Interestingly, a consistency check has been developed where Thiemann identity \eqref{Thiemann identity} is used \cite{giesel2006consistencyI,giesel2006consistencyII} to define a different flux operator, which is consistent with the usual one only for $\hat{V}_{v,AL}$ when $c_{AL}=l_p^3/\sqrt{48}$, but not for $\hat{V}_{v,RS}$, due precisely to factors $\hat{\epsilon}(e_i,e_j,e_k)$. Finally, some properties of the volume operator $\hat{V}_{v,AL}$ (that we will refer from now on as the volume operator) are \cite{Perez:2004hj}
\begin{itemize}
\item Due to the Gauss constraint, we know that for each vertex the total angular momentum is 0. Therefore, for a 3-valent vertex, we may write one flux as a linear combination of the other two, and then due to the presence of $\epsilon^{abc}$ in \eqref{classical regulated volume operator} makes $Q_{C^\epsilon_I}=0$. So the volume operator vanishes on 3-valent vertices.
\item The action of $\hat{V}_{v,AL}$ on planar edges vanishes for the same reasons.
\item The spectrum of $\hat{V}_{v,AL}$ is discrete. 
\end{itemize}

\subsection{The length operator}

The classical expression for the length of a curve $c: [0,1]\rightarrow \Sigma$ is
\begin{equation}
l(c)=\int_0^1\sqrt{q_{ab}(c(t))\dot{c}^a(t)\dot{c}^b(t)} dt=\int_0^1\sqrt{e^i_a(c(t))e^j_b(c(t))\dot{c}^a(t)\dot{c}^b(t)\delta_{ij}} dt.
\end{equation}
If we express the metric in terms of Ashtekar variables \cite{Pullin:2017}
\begin{equation}
q_{ab}=\frac{k}{4}\epsilon_{acd}\epsilon\indices{_{b}^{ef}}\epsilon^{ijk}\epsilon_{imn}\frac{P^c_j P^d_k P^m_e P^n_f}{\det P},
\end{equation}
we can see that the expression is non polynomial in the electric flux and therefore we cannot find a regularization similar to those used for the area and volume \cite{Pullin:2017}. Instead of that, different possible length operators have been proposed using the Thiemann identity \eqref{Thiemann identity} \cite{thiemann1998length}, Tikhonov regularization for the inverse $\hat{V}_{v,RS}$ operator \cite{bianchi2009length}, or in terms of other geometrical operators and $\hat{V}_{v,AL}$ \cite{ma2010new}.

\section{Diffeomorphism constraint}
The aim of this section is explaining how to implement the action of the diffeomorphism constraint on the spin network. The usual procedure would consist on writing down the classical expression and trying to promote the flux and holonomy operators to quantum operators. Taking \eqref{classical diffeomorphism constraint}, we can see that it involves a curvature term, which in differential geometry is usually defined in terms of parallel transport along small loops. The problem is that, due to background independence, we are unable to distinguish the `size' of such loops when they are diffeomorphic to each other, so the procedure fails as it is not well defined. In other words, the generator of diffeomorphisms does not exists as operator. We must resort to a different method using finite diffeomorphisms. These transformations form a group and can be used in a procedure called group averaging techniques, to obtain $\mathcal{H}_{\text{Diff}}$.

We define a group averaging map $\eta$, as an antilinear map $\eta: \mathcal{D}\rightarrow  \mathcal{D}^\star$ from a dense domain $\mathcal{D}\subset \mathcal{H}$, to the space  $\mathcal{D}^\star$ of complex linear mappings on $\mathcal{D}$, called \textit{algebraic dual}. This map must be invariant under the action of an unitary representation of a group G on $\mathcal{H}$, and must fulfil \cite{Pullin:2017}

\begin{enumerate}
\item $\forall \psi_1 \in \mathcal{D}, \eta(\psi_1)\in  \mathcal{D}^\star$ is invariant under the action of the representation of  an element of G, $\hat{U}(a)$:
\begin{equation}
 \quad \eta(\psi_1)[\hat{U}(a)\psi_2]=\eta(\psi_1)[\psi_2] \quad \forall a \in G, \psi_2\in \mathcal{D}
\label{property 1 of eta}
\end{equation}
\item $\eta$ is positive and real:
\begin{equation}
\eta(\psi_1)[\psi_2]=\overline{\eta(\psi_2)[\psi_1]}, \quad \eta(\psi_1)[\psi_1]\geq 0, \forall \psi_1,\psi_2\in \mathcal{D}
\end{equation}
\item $\eta$ commutes with the strong observables $\hat{O}\in \mathcal{O}$:
\begin{align}
\eta(\psi_1)[\hat{O}\psi_2]=\eta(\hat{O}^\dagger\psi)[\psi_2],\quad \forall \psi_1,\psi_2\in \mathcal{D}, \forall \hat{O}\in \mathcal{O}.\label{property 3 of eta}\\
\hat{O}\in \mathcal{O}\iff \hat{O}, \hat{O}^\dagger : \mathcal{D}\rightarrow \mathcal{D}, \quad \hat{U}(a)\hat{O}=\hat{O}\hat{U}(a) \quad\forall a\in G.\label{strong observables}
\end{align}
\end{enumerate}
Also, we define the action of $\hat{O}$ on $\Psi \in  \mathcal{D}^\star$ as $\hat{O}\Psi(\psi):=\Psi(\hat{O}\psi)$, and an inner product on  $\mathcal{D}^\star$ as $\braket{\eta(\psi_1),\eta(\psi_2)}_G:= \eta(\psi_2)[\psi_1]$. Finally, it is worth mentioning that the mathematical structure that we are using is $\mathcal{D}\subset \mathcal{H}\subset \mathcal{D}^\star$, and it is usually called second rigged\footnote{There is also the space of continuous antilinear functionals of $\mathcal{D}$, denoted as $\mathcal{D}^\times$, and the space of continuous linear functionals of $\mathcal{D}$,   $\mathcal{D}'$. In standard quantum mechanics, kets live in $ \mathcal{D}'$, and bras in $\mathcal{D}^\times$. $\mathcal{D}^\star$ (not necessarily continuous functionals) is endowed with the *-weak topology of point-wise convergence of nets.} Hilbert space or Gel'fand triple \cite{Giesel:2012ws}.

If there existed a measure $da$ for the group of diffeomorphisms, we could use it to define a candidate for the group averaging map $\eta(\ket{\psi}):=\int_G da \bra{\psi}\hat{U}(a)$. However, for $G=$Diff, no such measure is known so we must do something else. Let $[s]$ be the set of diffeomorphic, distinct (and therefore orthonormal with respect to the Ashtekar-Lewandowski measure) spin network functions $\ket{s}$.
Let us then formally propose 
\begin{equation}
\eta(\ket{s})=\eta_{[s]}\sum_{\ket{\overline{s}}\in[s]}\bra{\overline{s}}, \quad \eta_{[s]}\in \mathbb{R}^+.
\end{equation}
However, even if this is a formal expression, its application $\eta(\ket{s})\ket{s_2}$ will either result in $\eta_{[s]}$ if $\ket{s_2}\in[s]$, or $0$ otherwise, since the inner product \eqref{scalar product} will vanish always for any spin network not identically equal to $\ket{s_2}$. It is also clear that property \eqref{property 1 of eta} holds. Define $\text{Sym}_s\subset \text{Diff}$ as the subset that leaves $s$ invariant. Then
\begin{equation}
\eta(\ket{s}):=\eta_{[s]}\sum_{\phi \in \text{Diff}/\text{Sym}_s}(\hat{U}(\phi)\ket{s})^\dagger.
\label{action of eta}
\end{equation}
In \cite{Pullin:2017} it is shown that we can define the constant $\eta_{[s]}$ uniquely up to a global constant. Then, for $\forall \psi_{\alpha,f},\tilde{\psi}_{\tilde{\alpha},\tilde{f}}\in \mathcal{H}_{kin}$ define the diffeomorphism invariant inner product as
\begin{equation}
([\psi_{\alpha,f}]|\coloneqq \eta(\psi_{\alpha,f})\quad \Rightarrow \braket{[\psi_{\alpha,f}],[\tilde{\psi}_{\tilde{\alpha},\tilde{f}}]}_{Diff}=([\psi_{\alpha,f}]\ket{\tilde{\psi}_{\tilde{\alpha},\tilde{f}}}= \sum_{\phi \in \text{Diff}/\text{Sym}_\psi}\bra{\psi_{\alpha,f}}\hat{U}^\dagger(\phi)\ket{\tilde{\psi}_{\tilde{\alpha},\tilde{f}}}.
\end{equation}
This inner product defines $\mathcal{H}_{\text{Diff}}$.

\section{Hamiltonian constraint and Thiemann's work}\label{section:Hamiltonian constraint}
In this section we aim to explain what was Thiemann's original proposal to solve the Hamiltonian constraint. Although there are issues with the approach, it is worth understanding the original work, as it is the basis for more modern research.

We start by the expressions for a  3-volume $V$, and the integrated extrinsic curvature
\begin{equation}
V=\int_{\Sigma}d^3x\sqrt{det(q)}, \quad \bar{K}:=\int_{\Sigma}d^3xK^i_aE^i_a,
\label{Classic_volume/curvature}
\end{equation}
where $K^i_a=\gamma^{-1}(A^i_a-\Gamma^i_a)$ (see \eqref{connection}). Recall also the Hamiltonian constraint
\begin{equation}
\mathcal{C}=\underbrace{\frac{k\gamma^2}{2}\frac{\epsilon\indices{^{mn}_j}P_m^aP_n^b}{\sqrt{det(q)}}(F^j_{ab}}_{\mathcal{C}_E: \text{ Euclidean constraint,  } }-(1+\gamma^2)\epsilon^{jkl}K^k_aK^m_b )=:\mathcal{C}_E+\mathcal{T}.
\label{Hamiltonian_constraint}
\end{equation}
Thiemann's strategy consisted on first expressing the Hamiltonian constraint in terms of Poisson brackets, which we will promote to operators, and  then expressing the curvature $F^j_{ab}$ in terms of holonomies. The first step is  writing the triads like \cite{Ashtekar:2004eh}:
\begin{equation}
e_a^i=\frac{(k\gamma)^2}{2}\epsilon_{abc}\epsilon^{ijk}\frac{P^b_j P^c_k}{\sqrt{det(q)}},
\end{equation}
and also in terms of a Poisson bracket \cite{Ashtekar:2004eh}
\begin{equation}
e_a^i(x)=\frac{2}{k\gamma}\{A^i_a(x),V\}.
\end{equation}
So, equalling these two expressions we get (notice the change from $P$ to $E$)
\begin{equation}
\epsilon_{abc}\frac{E_i^a E_j^b\epsilon^{ijk}}{\sqrt{det(q)}}(x)=\frac{4}{k\gamma}\{A_c^k(x),V\},
\label{Thiemann identity}
\end{equation}
where we used \eqref{normalized E}. Finally, we use this identity to write the smeared $\mathcal{C}_E(N)$ as
\begin{equation}
\mathcal{C}_E(N)=\frac{2}{k^2\gamma}\int_{\Sigma}d^3xN\epsilon^{abc}\delta_{ij} F_{ab}^i\{A_{c}^j,V\}=\frac{2}{k^2\gamma}\int_{\Sigma}d^3xN\epsilon^{abc}tr(F_{ab}\{A_{c},V\}).
\label{Thiemann_euclidean_hamiltonian_constraint}
\end{equation}
For the rest of the Hamiltonian constraint, first 
 we have, using \eqref{connection}, \cite{Giesel:2012ws}
\begin{equation}
K^j_a=\gamma^{-1}(A^i_a-\Gamma^i_a)=\frac{1}{k\gamma}\{A^j_a,\bar{K}\}.
\end{equation}
We must also express $\bar{K}$ as a commutator, more precisely
\begin{equation}
\bar{K}=\gamma^{-3/2}\{\mathcal{C}_E(1),V\},
\end{equation}
so that
\begin{equation}
\mathcal{T}(N)=\frac{2}{k^4\gamma^3}\int_{\Sigma}d^3xN\epsilon^{abc}\epsilon_{ijk} tr(\{A_a^i,\bar{K}\}\{A_b^j,\bar{K}\}\{A_c^k,V\}),
\end{equation}
We substitute the Poisson brackets by commutators, and also write the extrinsic curvature in terms of the holonomy (since the extrinsic curvature does not have a well defined quantum operator but the holonomy does).
At an edge $e$ and small loop $\alpha$ of sizes $\epsilon$ and $\epsilon^2$ respectively, we may infinitesimally expand the holonomy as
\begin{align}
h_{e_a}[A]&=1-\frac{i}{2}\sigma_iA^i_a\epsilon+O(\epsilon^2)\\
%
%
h_{\alpha_{ab}}[A]&=1-\frac{i}{2}\sigma_iF^i_{ab}\epsilon^2+O(\epsilon^3),
\end{align}
where $\dot{e}^a$ is the tangent unit vector to $e^a$, and we have used the fundamental representation of $SU(2)$ with generators $\tau_i= i\sigma_i/2$. Therefore, infinitesimally \cite{Perez:2004hj}
\begin{align}
h_{\alpha_{ab}}[A]-h_{\alpha_{ab}}^{-1}[A]=2\epsilon^2F^i_{ab}\tau_i+O(\epsilon^4),\\
h_{e_a}^{-1}[A]\{h_{e_a}[A],V\}=\epsilon\{A^i_a,V\}+O(\epsilon^2).
\end{align}
The procedure now consists on
\begin{enumerate}
\item First we perform a triangulation of $\Sigma$
\item To each cell $\Delta$ associate edges ($O(\epsilon)$) and loops ($O(\epsilon^2)$), and fix one point $v(\Delta)$
\item The loops $\alpha_i(\Delta)$ and edges $e_i(\Delta)$ have tangents that span the tangent space at the vertex $v(\Delta)$.
\item Using $d^3x\approx\epsilon^3$, we approximate $\mathcal{C}_E(N)$, \eqref{Thiemann_euclidean_hamiltonian_constraint}, by
\begin{equation}
\begin{split}
\mathcal{C}_E(N)\approx\mathcal{C}_E^{(\Delta)}&:= \frac{2}{k^2\gamma}\sum_{\Delta}N(v_\Delta)\sum_{i=1}^3\epsilon^{abc} tr(F_{ab}\{A_c,V\})\epsilon^3\\
&=\frac{4}{k^2\gamma}\sum_{\Delta}N(v_\Delta)\sum_{i=1}^3[(h_{\alpha_i(\Delta)}^{-1}-h_{\alpha_i(\Delta)})h_{e_i(\Delta)}^{-1}\{h_{e_i(\Delta)},V\}].
\label{Classical_hamiltonian_constraint}
\end{split}
\end{equation}
\item A similar procedure can be carried out for $\mathcal{T}$ substituting the connections by holonomies and using the identity  $\hat{K}=\{V,\mathcal{C}_E\}$ \cite{Giesel:2012ws}.
\end{enumerate}
In order to quantize the expression \eqref{Classical_hamiltonian_constraint} we substitute the Poisson brackets for commutators, and promote the expressions to operators
\begin{equation}
\hat{C}_E^{(\Delta)}(N)=\frac{4}{k^2\gamma}\sum_{\Delta}N(v_\Delta)\sum_{i=1}^3[(h_{\alpha_i(\Delta)}^{-1}-h_{\alpha_i(\Delta)})h_{e_i(\Delta)}^{-1}[h_{e_i(\Delta)},\hat{V}]]
\label{Quantum_hamiltonian_constraint}
\end{equation}
as well as the corresponding for $\mathcal{T}$. Note that the explicit dependence on $\epsilon$ has disappeared and now it is only implicit in the triangulation. After this, the idea is to take this triangulation (cells can be of arbitrary shape) to the continuum limit, and see if it converges. 
Qualitatively one may say the following \cite{Giesel:2012ws}:
\begin{itemize}
\item Since the volume operator acts only on the vertices, the same happens for the Hamiltonian constraint operators.
\item It acts on the spin network by creating and annihilating loops and edges (new edges are called exceptional) at the vertices. This creates new nodes which are invisible to the Hamiltonian constraint and carry no volume (since 3-valent nodes are annihilated by the volume operator).
\end{itemize}
Due to problems related to convergence when $\epsilon \rightarrow 0$, Thiemann imposed restrictions about how the triangularization should be carried out. Based on the diffeomorphism invariance of $\hat{C}^{(\Delta)}(N)$, he requested that:
\begin{itemize}
\item For any of the vertex of the spin network, contributions for two given values of $\epsilon$ should be diffeomorphic (called regulator covariance).
\item $\exists \epsilon'>0$ such that for any spin network and  $\forall \epsilon < \epsilon'$, $\hat{C}^{(\Delta)}(N)$ is well defined (called uniform regulator covariance), and this will happen when there is at most one vertex per cell.
\end{itemize}
These two properties allow to see that, for any diffeomorphism invariant state $([\phi]| \in Cyl^{*}\subset \mathcal{H}_{Diff}$ we have that $ (\phi|\hat{C}^{(\Delta)}_\epsilon(N)\ket{\psi}$ is well defined and independent of $\epsilon$:
\begin{equation}
(\phi|\hat{C}^{(\Delta)}(N)\ket{\psi}=\lim_{\epsilon\rightarrow 0} (\phi|\hat{C}^{(\Delta)}_\epsilon(N)\ket{\psi}.
\end{equation}
This does not however mean that the operator $\hat{C}^{(\Delta)}(N)$ is well defined. In fact, we may only say that \cite{Pullin:2017}
\begin{equation}
\lim_{\epsilon\rightarrow 0}\hat{C}_\epsilon(N)=\hat{C}_{\epsilon_0}(N), \forall \epsilon_0<\epsilon'.
\end{equation}
This means that the limit is not unique and depends on the choice of $\epsilon_0$. 

The derivation that we have performed displays ambiguities and choices like the operator ordering or the choice of the representation, which will lead to several problems. In the last chapter we shall see more on these issues, and explore a more modern proposal called Master constraint, that aims to at least partially solve some of those problems the Hamiltonian operator has. 

\section{Conclusions}

In this chapter we have faced the issue of quantizing and solving the three constraints. First we have seen how restricting our spin network functions to those that are $SU(2)$ gauge invariant provides a solution to the Gauss constraint. 

Then we have made a digression to talk about the kinematical geometrical operators which will be necessary to explain the entropy of black holes in the case of the area operator, or solving the Hamiltonian constraint for the volume operator. We have briefly mentioned the length operator too.

Finally, we have seen how group averaging techniques provide a solution to the diffeomorphism constraint, and reviewed the original work of Thiemann in the Hamiltonian constraint that, although with several issues and seemly arbitrary choices, will be fundamental to understand more modern techniques used to solve the Hamiltonian constraint (which we will explore a bit on the last chapter).

\chapter{Spin foams}\label{ch:spinfoams}

Formally one may write the solution of the scalar Hamiltonian constraint as the kernel of the of a projector from $\mathcal{H}_{kin}$ to $\mathcal{H}_{phys}$
\begin{equation}
P=\int \mathcal{D}[N] e^{i\int_{\Sigma}N(x)\hat{C}(x)},
\end{equation}
where in the exponent we can recognize the smeared Hamiltonian constraint. This resembles the path integral formulation of Quantum Field Theory, and enables us to define a physical product:
\begin{equation}
\braket{s,s'}_{p}:=\braket{Ps,s'}
\end{equation}
for $s, s'\in \mathcal{H}_{kin}$.
Therefore, one may attempt constructing a path integral approach for LQG. This research line started in 1968, when Ponzano and Regge proposed the first path integral formulation of a spin foam model for (Euclidean, 3D) gravity \cite{regge1961general,regge2000discrete}:
\begin{equation}
Z=\int\mathcal{D}g_{\mu\nu}e^{\frac{i}{\hbar}S[g_{\mu\nu}]}; \quad S[g_{\mu\nu}]=\frac{c^4}{8\pi\hbar G}\int_\mathcal{R}d^3x\sqrt{g}R+\frac{c^4}{8\pi\hbar G}\int_{\partial\mathcal{R}}d^2x\sqrt{h}K
\label{Regge and Ponzano}
\end{equation}
where the action is defined as the Hilbert-Einstein action for a compact region $\mathcal{R}$ with the induced geometry of a flat tetrahedron \cite{Pullin:2017}, $S[\eta_{\mu\nu}]$

The analogous path integral formulation for 4D Lorentzian General Relativity was not found until recently (see \cite{Engle2008,Freidel:2007py,Pereira2008,Engle:2007wy}). 
To review this model (usually called EPLR model\footnote{Actually there are two very closely related models: the EPRL model after Engle-Pereira-Rovelli-Livine, and the FK model, after Freidel-Krasnov. We shall review here the basics of them.}) we shall start our description from  a topological field theory with some defects that unfreeze a finite number of degrees of freedom. After that, we will see what are the variables we need to describe our space, how to derive the area operator, and finally study the transition amplitudes of our theory\footnote{We will work with $\Lambda=0$ to make things simpler, but the general case has also been worked out \cite{Han:2010pz,Fairbairn:2010cp,Han:2011aa,Haggard:2014xoa}.}.
\section{A topological theory of gravity}
A topological theory is a theory with no local degrees of freedom. This is the case for general relativity on 2+1 dimensions. Consider now a 4 dimensional manifold with topology $M=\Sigma \times \mathbb{R}$, being $\Sigma$ compact. We define a topological theory of BF type, as one with gauge group $SO(3,1)$ and action
\begin{equation}
S_{top}=\int_M B_{IJ}\wedge F^{IJ}; \quad I,J\in\{0,...,3\}
\label{BF action}
\end{equation}
where we have a Lorentz connection $\omega^{IJ}=\omega^{IJ}_\mu(x)dx^\mu$ with curvature $F^{IJ}:=d\omega^{IJ}+\omega\indices{^I_K}\wedge \omega^{KJ}$ and $B^{IJ}:=B^{IJ}_{\mu\nu}(x)dx^\mu\wedge dx^\nu$. This action is invariant under local Lorentz transformations and Diff$(M)$. There is one more local symmetry, namely the shift of $B$ as $B^{IJ}\rightarrow B^{IJ}+d\Lambda^{IJ}+\omega\indices{^I_K}\wedge\Lambda^{KJ}+\omega\indices{^J_K}\wedge\Lambda^{KI}$ for $\Lambda^{IJ}$ a one-form. The equations of motion of the theory are 
\begin{equation}
F=0, \qquad dB^{IJ}+\omega\indices{^I_K}\wedge B^{KJ}+\omega\indices{^J_K}\wedge B^{KI}=0.
\label{EoM_topological}
\end{equation}
The first equation indicates that locally there is no curvature, and the second how we can shift the $B$ field. Therefore, it is clear that the solution is unique except for a possible gauge transformation and shift.

Now let us introduce a co-frame field $e^I=e^I_\mu(x)dx^\mu$, and define de Hodge star as in \eqref{Hodge}, 
for the curvature and torsion ($T^{IJK}=de^I+\omega^I_J\wedge e^J$).
It is well known that we can rewrite the Hilbert-Einstein action to the form of the Palatini action \cite{Rovelli:2014ssa}: 
\begin{equation}
S[e,\omega]=\frac{c^4}{8\pi\hbar G}\int_M e^I\wedge e^J\wedge ^\star(F)_{IJ}=\frac{c^4}{8\pi\hbar G}\int_M \frac{1}{2}\epsilon_{IJKL}e^I\wedge e^J\wedge F^{IJ},
\label{Palatini action}
\end{equation}
where the tetrad and the connection are taken like independent variables. Now let us add a term dependent on $\gamma$ such that 
\begin{equation}
S[e,\omega]=\frac{c^4}{8\pi\hbar G}\int_M e^I\wedge e^J\wedge ^\star(F)_{IJ}-\frac{1}{\gamma}e_{I}\wedge e_{J}\wedge F^{KL}.
\label{Holst action}
\end{equation}
This is called the Holst action \cite{holst1996barbero}. If we vary this action with respect to the two variables, we obtain Einstein equations for pure gravity, and the torsionless condition:
\begin{equation}
e^{J}\wedge ^\star(F)_{IJ}=0, \qquad e^{I}\wedge(de^J+\omega^J_K\wedge e^K)=0.
\end{equation}
The important point here is that the $\gamma$ dependent term in the action does not modify the classical equations of motion. 
We can recover the topological action \eqref{BF action} from \eqref{Holst action} by freezing the local degrees of freedom, that is, setting
\begin{equation}
B^{IJ}=\frac{c^4}{8\pi\hbar G}\left(\frac{1}{2}\epsilon\indices{^{IJ}_{KL}}e^{K}\wedge e^{L}-\frac{1}{\gamma}e^{I}\wedge e^{J}\right),
\label{B field}
\end{equation}
and we can also invert this relation to obtain
\begin{equation}
e^I\wedge e^J=\frac{-8\pi\hbar G}{c^4}\frac{\gamma^2}{\gamma^2+1}\left(\frac{1}{2}\epsilon\indices{^{IJ}_{KL}}B^{KL}+\frac{1}{\gamma}B^{IJ}\right).
\label{inverse B field}
\end{equation}
This means that we can formulate general relativity as a topological theory \eqref{BF action}, requiring $B$ to be of the form \eqref{B field} for some $e^I$, what unfreezes degrees of freedom, or equivalently that the parenthesis in \eqref{inverse B field} is a simple 2-form $\Sigma^{IJ}$. This is called the \textit{simplicity constraint} and will be important in the quantization procedure.
\section{Discretization and variables.}
We define a spin foam model as a topological theory \eqref{BF action} where a finite number of degrees of freedom have been unfrozen by the introduction of a cellular decomposition. Although it is not necessary (see for instance \cite{Pullin:2017}) for simplicity we will consider that these cells have the form of tetrahedra. The set of (disjoint) cells ($\Delta$) in the 4 dimensional manifolds ($M$) is defined by points (p), segments (s), triangles (t) and tetrahedra ($\tau$). We also need to define a 2-complex ($\Delta^*$) as the dual of $\Delta$.
\begin{table}[htbp]

\begin{tabular}{|l|l|l|l|}
\hline
Cell $\Delta$ & Two-complex $\Delta^*$ & Boundary $\partial \Delta$ & Boundary graph $\partial \Delta^*=\Gamma$ \\ \hline
4-simplex ($v$) & Vertex(v) &  &  \\ \hline
Tetrahedron ($\tau$) & Edge (e) & Tetrahedron ($\tau$) & Node (n) \\ \hline
Triangle (t) & Face (f) & Triangle (t) & Link (l) \\ \hline
Segment (s) &  &  &  \\ \hline
Point (p) &  &  &  \\ \hline
\end{tabular}
\label{Cellular decomposition notation}
\caption{Terminology and notation of the cellular decomposition.}
\end{table}

\begin{figure}
\begin{center}
\includegraphics[width=0.4\textwidth]{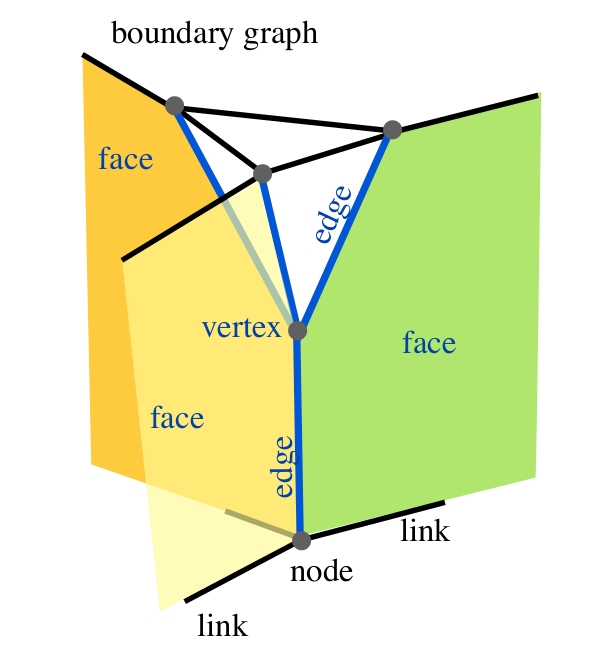}
\end{center}
\label{2-complex figure}
\caption{Graphical representation of the 2-complex. Figure taken from \cite{Rovelli:2014ssa}.}
\end{figure}

We want to ensure that the 2-skeleton of the cellular decomposition, $\Delta_2$ is everywhere space-like, and most importantly, unfreeze some degrees of freedom by imposing the simplicity constraint $\Sigma^{IJ}$. For that we introduce a timelike vector $t^I$, and define the action as 
\begin{equation}
S_{top}=\int_M B_{IJ}\wedge F^{IJ}+\int_{\Delta_2} \lambda_I t_J \Sigma^{IJ},
\label{EPLR-FK action}
\end{equation}
where $\lambda^I$ is a 0-form working as a Lagrange multiplier that ensures that $t_J \Sigma^{IJ}=0$. If we know define
\begin{equation}
B^I=B^{IJ}t_J, \quad E^I=\frac{1}{2}\epsilon^{IJKL}B_{JK}t_L, \Rightarrow  B^It_I=0=E^It_I \Rightarrow B^0=0=E^0
\label{B^I E^I}
\end{equation}
by antisymmetry of $B^{IJ}$ (see \eqref{B field}), we can rewrite the simplicity constraint as
\begin{equation}
B^I=\gamma E^I \text{ on } \Delta_2.
\label{simplicity constraint}
\end{equation}
Equivalently, for the electric and magnetic fluxes through a 2-cell $t_l$,
\begin{equation}
\vec{L}_l=\int_{t_l}\vec{E}, \quad \vec{K}_l=\int_{t_l}\vec{B} \quad \Rightarrow \quad \vec{K}_l=\gamma \vec{L}_l.
\label{simplicity constraint 2}
\end{equation}

\section{Variables and the group $SL(2,\mathbb{C})$.}
The variables that we will use for the discretization of the connection and tetrad are
\begin{equation}
\begin{aligned}
\omega \rightarrow U_e= \mathcal{P}\text{exp }i\int_e \omega \in SL(2,\mathbb{C})\\
e\rightarrow B_f^{IJ}=\int_{t_f}B^{IJ} \in sl(2,\mathbb{C}).
\end{aligned}
\label{discretization variables}
\end{equation}
The first equation indicates that instead of the connection we will use, as for canonical quantization, the holonomy along the edge $e$ of the 2-complex. It is simple to see that the holonomy is a group element (like in \eqref{holonomy}, except for that the group now is $SL(2,\mathbb{C})$ instead of $SU(2)$), and $B^{IJ}$ is in the group algebra, since it has two indices (it is a matrix). The reason to use $SL(2,\mathbb{C})$ and $sl(2,\mathbb{C})$ is that this group is the universal cover of 
$SO^\uparrow(3,1)$, used in Lorentzian general relativity. Let us recall briefly some properties of $SL(2,\mathbb{C})$.
The unitary representations $V^{(p,k)}$ of this group are infinite-dimensional, labelled by $p\in \mathbb{R}$ and $k\in \mathbb{Z}/2$, and generated by six hermitian operators $J^{IJ}=-J^{JI}$ where $J,I\in \{0,1,2,3\}$. Introducing a timelike vector $t^I$ we can define the generator of boosts and rotations, respectively
\begin{equation}
K^I=J^{IJ}t_J, \qquad L^I=\frac{1}{2}\epsilon \indices{^I_{JKL}}J^{KL}t^L.
\label{def K L}
\end{equation}
From this we know that $t_IK^I=0$, since the product of $t_It_J$ is symmetric in its indices but $J^{IJ}$ is antisymmetric. The same happens for the case of $t_IL^I=0$, with indices $I$ and $L$. This will mean that $K^I=(0,\vec{K}^i)$ and $L^I=(0,\vec{L}^i)$, and they will obey the usual commutation relations for generators of $SU(2)$. It is also important that \cite{Pullin:2017} 
\begin{equation}
V^{(p,k)}=\bigoplus_{j=k}^\infty V^{(j)},\quad V^{(j)}\text{ unitary repr. of } SU(2); \quad \dim(V^{(j)})=2j+1.
\label{SL(2,C) to SU(2) decomposition}
\end{equation}
 Every $V^{(p,k)}$ has two Casimirs \cite{Pullin:2017}
\begin{subequations}
\begin{align}
&C_1=|\vec{K}|^2-|\vec{L}|^2=\frac{1}{2}J_{IJ}J^{IJ}=p^2-k^2+1,\\
&C_2=\vec{K}\vec{L}=\frac{1}{8}\epsilon_{IJKL}J^{IJ}J^{KL}=pk.
\end{align}
\label{casimirs of SL(2,C)}
\end{subequations}
Using \eqref{simplicity constraint 2}, these two Casimirs  give
\begin{equation}
\begin{aligned}
&p^2-k^2+1=|\vec{K}|^2-|\vec{L}|^2=(\gamma^2-1)|L|^2=(\gamma^2-1)j(j+1)\\
&pk=\vec{K}\vec{L}=\gamma|\vec{L}|^2=\gamma j(j+1)\\
&j\rightarrow \infty \Rightarrow
p^2-k^2=(\gamma^2-1)j^2, \quad
pk=\gamma j^2
\Rightarrow
p=\gamma k = \gamma j
\end{aligned}
\end{equation}

Additionally, we can denote an orthonormal basis of these representations, using \eqref{SL(2,C) to SU(2) decomposition}, by $\ket{p,k;j,m}$. The previous equations allow us to define a map $Y_\gamma$
\begin{equation}
\begin{aligned}
Y_\gamma:& V^{(j)} \rightarrow V^{(\gamma j,j)}\\
&\ket{j,m}\rightarrow\ket{\gamma j,j;j,m}
\label{Y map}
\end{aligned}
\end{equation}
such that\footnote{Although our derivation only works approximately for large $j$, the relation \eqref{simplicity constraint 3} can be made exact for any $j$ using $p=\gamma (j+1),\text{ } k=j$. To see how, refer to page 168, \cite{Rovelli:2014ssa}.} for $j$ large, and for any $\psi, \phi \in \mathcal{H}_j$ being $\mathcal{H}_j$ a Hilbert space in $V^{(j)}$,
\begin{equation}
\bra{Y_\gamma \psi}\vec{K}-\gamma\vec{L}\ket{Y_\gamma \phi}=0.
\label{simplicity constraint 3}
\end{equation}

\subsection{The area operator}
Now we can do something interesting. Let us get the same result for the area operator\footnote{For the volume operator a similar derivation exists in the covariant formalism (see e.g. \cite{Rovelli:2014ssa}). It can be calculated from the formula 
\begin{equation}
\hat{V}_n=\frac{\sqrt{2}}{3}(8\pi G \hbar \gamma)^{\frac{3}{2}}\sqrt{|L_i\cdot (L_j\times L_k)|}.
\end{equation}} that we got in the canonical formalism, using this approach. Starting from \eqref{simplicity constraint}
\begin{equation}
E^I=\frac{1}{\gamma}B^I=\frac{1}{\gamma}B^{IJ}t_J=\frac{1}{\gamma}\frac{c^4}{8\pi\hbar G}\left(\frac{1}{2}\epsilon\indices{^{IJ}_{KL}}e^{K}\wedge e^{L}-\frac{1}{\gamma}e^{I}\wedge e^{J}\right)t_J.
\end{equation}
If we analyse this expression in the boundary, we realise that $(e^{I}\wedge e^{J}) t_J=0$ since $t_J$ is normal to $(e^{I}\wedge e^{J})$. Then we can drop the second term in the parenthesis. 
Defining the area and integrating we get, from \eqref{simplicity constraint 2},
\begin{equation}
\text{Area}_{t_l}=\frac{1}{2}\int\epsilon\indices{^{IJ}_{KL}}(e^{K}\wedge e^{L}) t_J \Rightarrow \vec{L}_l=\int_{t_l}\vec{E}=\frac{1}{\gamma}\int_{t_l}B^{IJ}t_J=\frac{1}{\gamma}\frac{c^4}{8\pi\hbar G}\text{Area}_{t_l},
\end{equation}
so that, as $\vec{L}\in SU(2)$ and the eigenvalues of $|\vec{L}|^2$ are $j(j+1)$,
\begin{equation}
\text{Area}=\frac{8\pi\hbar G \gamma}{c^4} \vec{L}_l=\frac{8\pi\hbar G \gamma}{c^4} \sqrt{j(j+1)}.
\label{covariant area}
\end{equation}
We then get that the area gap is \cite{Rovelli2004Quantum} $\Delta= 4\sqrt{3}\pi\hbar G \gamma c^{-3}\sim  10^{-66}\gamma\text{ cm}^2$. 
\section{Boundary Hilbert space.}
We have already introduced some variables on the surface, namely $\vec{K}$ and $\vec{L}$. What we want to do now is to construct a Hilbert space on a 3D boundary space (spin network), and see how it evolves calculating the transition amplitudes.

Recall that we are working now in the gauge group $SU(2)$, universal cover of $SO(3)\subset SO^\uparrow(3,1)$. The variables that we will use are the electric field $E^I$ defined in \eqref{B^I E^I}, and the connection $\omega$ restricted to the 2-cells of the decomposition, \cite{Pullin:2017} 
\begin{equation}
A_I=\omega_{IJ}t^J+\frac{\gamma}{2}\epsilon_{IJKL}\omega^{JK}t^L.
\end{equation}
They are conjugate variables, and we can recover the flux-holonomy algebra by defining the holonomy along a link l of $\Gamma=\partial \Delta^*$ \eqref{holonomy}, and the electric flux along faces,
%
%
\begin{eqnarray}
&h_l=\mathcal{P}\text{exp }i\int_l \vec{A}\cdot\frac{\vec{\sigma}}{2} \in SU(2); \quad h_e=\mathcal{P}\text{exp }i\int_e \vec{A}\cdot\frac{\vec{\sigma}}{2} \in SU(2)\\
&\vec{L}_l=\int_{t_l}\vec{E} \in su(2).
\end{eqnarray}
We can also see that $A_I t^I=0$ and $E_I t^I=0$ due the same argument that we applied for $t_IK^I=0$ and $t_IL^I=0$ in \eqref{def K L}.
By choosing an appropriate timelike $t^I$ we can get $A^I=(0,A^i)$ and $E^I=(0,E^i)$.

Now we want to construct the Hilbert space. Since our variables are assigned for each link, and they are in SU(2) (where as we saw already we have the invariant Haar measure), the first option would be $\mathcal{H}_\Gamma = L^2[SU(2)^L]$ where $L$ is the number of links of $\Gamma$, and $L^2$ indicates the square integrable functions. However, since we want to impose $SU(2)$ gauge invariance for each 3-cell (rotation of the frame in each cell), the physical Hilbert space of the theory will be
\begin{equation}
\mathcal{H}_\Gamma= L^2[SU(2)^L/SU(2)^N],
\end{equation}
for $N$ being the number of nodes and therefore of 3-cells. For this Hilbert space we can find an orthonormal basis, the spin network basis. The 2 elements needed are $SU(2)$ representation matrices, that we write as $D^{(j)}(h)\indices{^m_{m'}}=\bra{j,m}h\ket{j,m'}$, and intertwiners $\iota^{m_1,...,m_N}_{n;n_1...n_M}$. The discussion on spin network basis of chapter \ref{ch:kinematics} applies here.

It is also common to denote the spin network functions as $\psi_{j_l,i_n}(h_l)=\braket{h_l|\Gamma,j_l,i_n}$. Finally, we may use the $Y_\gamma$ map \eqref{Y map}, to assign to each function of the Hilbert space, a function of the Lorentz group via the identifications
\begin{eqnarray}
&D^{(j)}(h_l)\indices{^m_{m'}} \rightarrow D^{(\gamma(j+1),j)}(U_e)\indices{^{jm}_{jm'}}=\bra{j,m}Y^\dagger_\gamma U_e Y_\gamma\ket{j,m'}\\
&\psi_{j_l,i_n}^{\gamma}(U_e)=\left(\bigotimes_{l\in \Gamma} D^{(\gamma(j_l+1),j_l)}(U_e)\right)\cdot \left( \bigotimes_{n\in \Gamma} Y_\gamma \iota_n  \right),
\end{eqnarray}
where $Y_\gamma$ are defined using the action of $Y_\gamma$ on the spin network functions \eqref{SNF}.
\section{Transition amplitudes}
Up to now we have worked on the boundary space. We would like to calculate the transition amplitudes
\begin{equation}
W_\Delta (h_l):=\int dh_l \prod_f A_f(h_l),
\end{equation}
where $A_f$ indicates the face amplitude, yet to be defined. As we can see, this amplitude will depend on the cellular decomposition of the boundary $\Delta$. To derive the transition amplitude, we can start from the expression
\begin{equation}
W_\Delta(h_l)=\mathcal{N}\int dh_e \int dL_f e^{\frac{i}{8\pi \hbar G}\sum_fTr(h_f L_f)},
\end{equation}
where the exponent is the action on the cellular decomposition $\Delta$ and $h_f=\prod_{e\in f} h_e$ for all edges around a given face. Then, using $\delta(U)=\sum_j(2j+1)D^{(j)}(U)$, which is the equivalent for $SU(2)$ of $\delta(\phi)=(2\pi)^{-1}\sum_ne^{in\phi}$ for $U(1)$ \cite{Rovelli:2014ssa}, we can write
\begin{equation}
W_\Delta(h_l)=\mathcal{N}\int dh_e \prod_f \delta(h_f)=\mathcal{N}\int dh_e \prod_f\left(\sum_j(2j+1)D^{(j)}(U)\right).
\end{equation}
We can also expand $h_f=\prod_{e\in f} h_e$, and introduce variables $h_e=g_{ve}g_{ev'}$, with $h_{vf}=g_{ev}g_{ve'}$ ($v$ indicates a vertex, f a face and e and edge), and $g_{ve}=g^{-1}_{ev}$. So, \cite{Rovelli:2014ssa}
\begin{equation}
\begin{aligned}
W_\Delta(h_l)&=\mathcal{N}\int dg_{ve} \prod_f \delta(g_{ve}g_{ev'}g_{v'e'}g_{e'v''}...)\\
&=\mathcal{N}\int dh_{vf} dg_{ve} \prod_f \delta(g_{ve}g_{ev'}g_{v'e'}g_{e'v''}...)\prod_{vf}\delta(g_{e'v}g_{ve}h_{vf})\\
&=\mathcal{N}\int_{SU(2)} dh_{vf}\prod_f \delta(h_f) \prod_v A_v(h_{vf}),
\label{transition amplitude}
\end{aligned}
\end{equation}
where the vertex amplitude is 
\begin{equation}
A_v(h_{vf})=\int_{SU(2)} dg_{ve}\prod_f \delta(g_{e'v}g_{ve}h_{vf})= \sum_{j_f}\int_{SU(2)} dg'_{ve}\prod_{f}(2j_f+1)tr_{j_f}(g_{e'v}g_{ve}h_{vf}). 
\end{equation}
However, the boundary $\Gamma$ is also in $\Delta^*$ so we also need to impose $SL(2,\mathbb{C})$ invariance for the vertex amplitude. The rest of \eqref{transition amplitude} does not need any additional change as the kinematical space for 3d and 4d is the same, and the dynamics happen at the vertices (the Hamiltonian constraint only acts on vertices) \cite{Rovelli:2014ssa}. Using the map $Y_\gamma$ \eqref{Y map}:
\begin{equation}
A_v(h_{vf})=\sum_{j_f}\int_{SL(2,\mathbb{C})} dg'_{ve}\prod_{f}(2j_f+1)tr_{j_f}(Y^\dagger_\gamma g_{e'v}g_{ve}Y_\gamma h_{vf}),
\end{equation}
where 
\begin{equation}
tr_{j}[Y^\dagger_\gamma gg' Y_\gamma h]=tr_{j}[Y^\dagger_\gamma D^{(\gamma j, j)}(gg')Y_\gamma D^{(j)}(h)]=\sum_{mn}D^{(\gamma j, j)}_{jm,jn}(gg')D^{(j)}_{nm}(h).
\end{equation}
\section{Coherent states}
Let us make a brief introduction to coherent states, that will play an important role on finding the classical limit. Suppose we have a 3D system with spin. For example, if we take a tetrahedra in space,  for each face we have $L_i$ fulfilling that due to rotation invariance $\sum_i L_i=0$. We then define a coherent state one such that the dispersion $\Delta L_z=0$ for z arbitrary. Since the $L_i$ do not commute, we get the Heisenberg relations
$\Delta L_x \Delta L_y \geq \frac{1}{2}|\braket{L_z}|.
$
To saturate it, we choose the coherent state $\ket{j,j}$ so that $L_z\ket{j,j}=j\ket{j,j}$. Obviously, $\Delta L_z=0$ and $\braket{L_x}=0=\braket{L_y}$. So, it is easy to see that $\Delta L_x=\sqrt{j/2}=\Delta L_y$ and
\begin{equation}
\frac{\Delta L_x}{\sqrt{\braket{\vec{L}^2}}}=\frac{\sqrt{j/2}}{\sqrt{j(j+1)}}=\frac{1}{\sqrt{2(j+1)}}\stackrel{j\rightarrow \infty}{\rightarrow}0,
\end{equation}
meaning that the state becomes relatively well defined for large $j$. Needless to say that we can rotate the coherent state to point in an arbitrary direction: $\ket{j,\vec{n}}=U(R)\ket{j,j}=\sum_{m=-j}^j \phi_m(\vec{n})\ket{j,m}$, with $\phi_m(\vec{n})=\braket{j,m|U(R)|j,j}$. Finally, they provide a resolution of the identity
\begin{equation}
\mathbbm{1}_j=\frac{2j+1}{4\pi}\int_{S^2}d^2\vec{n}\ket{j,\vec{n}}\bra{j,\vec{n}}.
\end{equation}
Once we have coherent states, we can define coherent intertwinners as the rotational-invariant tensor product of those states \cite{livine2007new}
\begin{equation}
\begin{split}
\ket{\iota(\vec{n}_a)}=\sum_m \Phi_{m_1...m_N}(\vec{n}_a) \ket{j_1,m_1}...\ket{j_N,m_N};\\
\Phi_{m_1...m_N}(\vec{n}_a)=\int_{SU(2)}\prod_{a=1}^N\braket{j_a,m_a|U(h)|j_a,\vec{n}_a}.
\end{split}
\end{equation}
We are now able to describe any spin foam in terms of these coherent states and intertwinners. Then, these intertwinners can be used to rewrite the vertex amplitude and it is shown \cite{Pullin:2017,Rovelli:2014ssa} that from such vertex amplitude the 4D Lorentzian equivalent of \eqref{Regge and Ponzano} emerges. Therefore, coherent states are key to the study of the semiclassical limit of the theory.

This completes the formal statement of the theory, since we have formulated the dynamics. Some properties of the theory are \cite{Rovelli:2014ssa} that it fulfils the superposition principle, the locality principle, it is Lorentz invariant, is ultraviolet finite, with a positive cosmological constant it becomes IR finite too \cite{Pullin:2017}. The theory will reproduce general relativity when we take the classical limit \cite{Rovelli:2014ssa}.

If we go back to the beginning of the chapter though, it will be apparent that we were only taking into account a finite number of degrees of freedom. How can we take the limit such that we reproduce the infinite degrees of freedom of General Relativity? Note that since the theory is diffeomorphism invariant it does not make sense to speak of the smallness of a certain cellular decomposition. The only sensible thing we can say is that a graph $\Gamma$ will be more refined than another $\Gamma'$ if $\Gamma' \subset \Gamma$, and this will imply that $\mathcal{H}_{\Gamma'}\subset \mathcal{H}_{\Gamma}$ \cite{Rovelli:2014ssa}. We may also go from a more refined graph to a smaller one by setting some $j=0$. The inverse procedure will also allow us to take the continuum limit for the transition amplitudes.

Finally, let us make a remark. The procedure we have followed encounters two distinct kinds of discreteness: on one hand we have classically discretise the space, by a cellular decomposition. However, there is a second discreteness that comes directly from quantum mechanics, and that is the discreteness of areas and volumes. In the same line, it is important to distinguish between two limits: the continuum limit where we take a very refined graph $\Gamma$, and the classical limit where the spins of the network are large (and so are the area, volume...). To derive this classical limit, it will be necessary to use techniques such as saddle point, and sum over spin foams using group field theory techniques. Interested readers can find information in \cite{Rovelli:2014ssa}, for saddle point techniques, and \cite{Pullin:2017} for an introduction to group field theory techniques.

\section{A simple derivation of the black hole entropy}

Let us end our description of this theory by applying it to calculate the entropy of a Schwarzschild black hole. Our aim is to derive Steinbeck-Hawking formula for the entropy of a black hole of area $A$ \cite{Hawking:1974sw,Hawking:1975bhe}
\begin{equation}
S_{BH}=\frac{kc^3}{4\hbar G}A.
\label{hawking entropy}
\end{equation}
One begins by realising that if by definition the mass $M$ is the energy of a black hole measured at infinite, then, at a small distance $r$ from the event horizon with Schwarzschild radius $R=2G M$, we have to redshift its value
\begin{equation}
E(R+r)=\sqrt{g^{00}(R+r)}M=\frac{M}{\sqrt{1-\frac{2G M}{2G M+r}}}=M\sqrt{\frac{2GM+r}{r}}\approx M\sqrt{\frac{2G M}{r}},
\end{equation}
and near the horizon, the physical distance $d$ and coordinate distance $r$ are related by
\begin{equation}
d=\sqrt{g_{rr}}r=\frac{r}{\sqrt{1-\frac{2G M}{2G M+r}}}\approx r\sqrt{\frac{2G M}{r}}=\sqrt{2G M r}.
\end{equation}
Combining the two previous equations one can see that the energy measured at a small distance d from the events horizon $d$ is
\begin{equation}
E(d)=\frac{2G M^2}{d}.
\label{black hole energy at d}
\end{equation}
We know recall that the area of the black hole is that of a sphere of radius $R$, $A=4\pi(2GM)^2$, and we can calculate that the acceleration measured by an observer at a fixed distance $d$ from the events horizon is $a=1/d$\footnote{To derive it start from setting $A^a=\frac{D}{Ds}V^a=V^b\nabla_b V^a$ for $V^a=k \partial_t $. Then normalize $V$ such that $g(V,V)=-1$ and the proper acceleration will be given by $a=\sqrt{g(A,A)}=\sqrt{g_{ab}A^aA^b}$. This is what is done in ex. 4 ps 4, GR I in the MSc in Mathematical and Theoretical Physics in Oxford.} \cite{Rovelli:2014ssa}. Substituting in \eqref{black hole energy at d}, we get the expression derived by Frodden, Ghosh and Perez\footnote{Alternatively \cite{bianchi2012entropy}, consider that as the observer is moving in an uniformly accelerated way, the generator of the proper time evolution (Hamiltonian) is $H=a\hbar K_z$. For the observer, the event horizon is locally flat with normal direction z, so can be represented by coherent state $\ket{j,j}$. It evolves according to $Y_\gamma: \ket{j,j} \mapsto \ket{\gamma j,j;j,j}$. Then, by \eqref{simplicity constraint 3} and \eqref{covariant area}, for $j$ large the energy is 
\begin{equation}
E=\bra{\gamma j,j;j,j}a \hbar K_z \ket{\gamma j,j;j,j}= \bra{\gamma j,j;j,j}a \hbar \gamma L_z \ket{\gamma j,j;j,j}=a \hbar \gamma j=a \hbar \gamma\frac{A}{8\pi G \hbar \gamma}= \frac{aA}{8\pi G}.
\end{equation}
The $Y_\gamma$ map is playing the role of Einstein equations (and Schwarzschild metric) here!} \cite{Ghosh:2011fc}
\begin{equation}
E=\frac{aA}{8\pi G}.
\label{FGP energy}
\end{equation}
We also need the fact, discovered by Unruh \cite{unruh1976}, that an observer with acceleration $a$ in vacuum will measure thermal radiation with temperature
\begin{equation}
T=\frac{\hbar a}{2\pi k},
\label{T Unruh}
\end{equation}
and taking the Clausius definition of entropy, we recover the Hawking formula
\begin{equation}
dS=\frac{dE}{T}=\frac{2\pi k}{\hbar a}\frac{a c^3 dA}{8\pi G}=\frac{k c^3}{4\hbar G}dA.
\end{equation}
Now we will use what we have learned of the operator of area, to calculate $S_{BH}$ from a kinematical point of view. We will calculate the entropy by counting states with different energies. At temperature $T=\beta^{-1}$, the probability $p_n$ of a system of being at a certain energy $E_n$ fulfils $p_n \propto e^{-\beta E_n/\hbar}$.
On the other hand, the area contribution of one link with spin $j$ at the horizon is given by \eqref{covariant area}. Since in this link we have a $SU(2)$ irreducible representation of spin $j$, there are $(2j+1)$ orthogonal states. Then
\begin{equation}
p_j(\beta) \propto (2j+1)e^{-\beta E_n/\hbar}=(2j+1)\exp{\left(-\beta \frac{aA}{8\pi G}\right)}=(2j+1)e^{-\beta a\gamma\sqrt{j(j+1)}},
\end{equation}
where we have substituted \eqref{FGP energy} for the energy and \eqref{covariant area} for the area. We fix the proportionality constant imposing $\sum_j p_j(\beta)=1$. 

The final expression is then
\begin{equation}
p_j(\beta)=Z^{-1}(\beta)(2j+1)e^{-\beta a\gamma\sqrt{j(j+1)}},\quad Z(\beta)=\sum_j (2j+1)e^{-\beta a\gamma\sqrt{j(j+1)}}.
\end{equation}
We will need the thermodynamical equations 
\begin{equation}
S=-\sum_j p_j \ln(p_j) \quad\Rightarrow\quad
%
%
E=TS+F;\qquad
F:= -T\ln Z.
\end{equation}
Finally, we need to assume that the state is at the Unruh temperature \eqref{T Unruh},
\begin{equation}
S=\beta(E-F)=\beta E+ \ln Z = \frac{2\pi k}{a\hbar}\frac{aA}{8\pi G}+ \log Z=S_{BH}+\log \sum_j (2j+1)e^{-\beta a\gamma\sqrt{j(j+1)}}.
\end{equation}
If $\log Z$ vanishes, we recover $S_{BH}$. This happens (numerical calculation) for $\gamma_0=0.274...$ This result has been derived for the Schwarzschild and Kerr black hole, and also using dynamical aspects of the theory (see \cite{Rovelli:2014ssa}), where instead of focusing on fluctuations of the event horizon on considers entanglement entropy of particles across the horizon.
%
%

\section{Conclusions}

In this chapter we have reviewed a different research approach to LQG: spin foams. We started this path integral formalism from a topological theory with a simplicity constraint that unfreezes some degrees of freedom. Then we have discretised our space in cells and we have calculated the area  operator. After that, with the help of boundary variables, we have defined our transition amplitudes and made a short introduction to coherent states, needed to calculate the classical limit of GR. Finally, we have derived the LQG calculation for the Schwarzschild black hole entropy.
\chapter{Issues and open questions}\label{ch:Issues}
As we have seen so far, Loop Quantum Gravity is not yet a finished theory. It has some open issues to tackle and it also happens that the severity of these problems are not rated equally by everyone. The aim of this chapter is to review different points of view of what problems need to be address, and also explain some of the choices we have made in the previous description of the theory. In particular I would like to explore some of the criticism expressed in papers \cite{nicolai2005loop,nicolai2007loop} and what is the point of view of the LQG community on such issues.

\section{Problems with the Hamiltonian constraint}
As we saw in section \ref{section:Hamiltonian constraint}, Thiemann's work on the Hamiltonian constraint did not fully solved the theory, as there are yet some problems arising from this approach. 
Following \cite{nicolai2005loop} let us list the main problems with the proposed solution.
\begin{enumerate}
\item \textit{Ambiguities.} Let us list all the different ambiguities involved, following \cite{Thiemann:2006cf}\footnote{I will not discuss habitat ambiguities due to their technical complexity and the fact that \cite{Thiemann:2006cf} indicates that `[...] is not a matter of debate, the habitat construction presented in \cite{nicolai2005loop} is outdated. Habitats are unphysical and completely irrelevant in LQG'.}

\begin{enumerate}
\item \textit{Factor ordering ambiguities.} In \eqref{Quantum_hamiltonian_constraint} one can see that we have arbitrarily chosen to write terms depending on the honolomy ($h_{e_i(\Delta)}, h_{\alpha_i(\Delta)}$) to the left of those dependent on the flux ($\hat{V}$). In \cite{Thiemann:2006cf} it is argued that it is not possible to choose any other ordering, since the result would not be densely defined \cite{thiemann1998QSD}, because the resulting state of applying this other choice for the Hamiltonian constraint would be a linear combination of states with graphs directly dependent on the chosen triangularization $\Delta$. It would therefore not be normalizable in the continuum limit, so our choice is singled out. 
\item \textit{Election of the representation.} It is common to choose the $SU(2)$ representation $j=1/2$ for the holonomies $h_{e_i(\Delta)}, h_{\alpha_i(\Delta)}$, since it is the easiest choice. However, other values of $j$ are also possible, and lead to a different operator $\hat{C}^{(\Delta)}(N)$ with the same classical limit \cite{gaul2001generalized}.

Nevertheless, \cite{Thiemann:2006cf} gives two reasons for this choice. The first one is that in 3D (completely solved) higher spin representations leads to spurious solutions of the spectrum of $\mathcal{H}$ \cite{perez2006regularization}, and so it is expected in 4D.
Moreover, this ambiguity is also present in standard QFT, where the spectrum changes if one substitutes the momentum $\pi$ by $[F\pi F^{-1}+\overline{F}^{-1}\pi \overline{F}]/2$, for $F$ a functional of $\phi$. This is not usually done in QFT because it destroys the polynomiality of the Hamiltonian, so one uses a naturalness argument to choose $\pi$ as the conjugate variable. Since in LQG the Hamiltonian is not polynomial anyway, choosing $j=1/2$ is only a matter of simplicity.
\item \textit{Ambiguities in loop alignment.} In the LQG community, it is the standard to choose $h_{e_i(\Delta)}$ and  $h_{\alpha_i(\Delta)}$ to align with three of the edges of a given vertex. However, one could in principle imagine a different choice, where the new edges are not aligned with the pre-existing ones, giving rise to the new plaquette being freely floating, only connected to the graph at the vertex. As it is discussed in \cite{nicolai2005loop}, this is possible, although commonly excluded by hand due to background independence \cite{thiemann2001introduction}. 

In this line, \cite{Thiemann:2006cf} gives 2 reasons for this choice. The first is that, if instead of working with $\{A^j_a(x),V(R_x)\}$ (right hand side of \eqref{Thiemann identity}) one chooses to work with the left hand side of \eqref{Thiemann identity}, he will get an expression like
\begin{equation}
\begin{split}
&\int_\sigma d^3x\frac{1}{\sum_{v'\in V(\gamma)}\delta(x,v')\hat{V}_v}\sum_{v\in V(\gamma)}\sum_{e_1\cap e_2 =v}\int_0^1 dt \dot{e}^a_1 (t) \delta(x,e_1(t))
\\
&\int_0^1 ds \dot{e}^a_2 (s)\delta(x,e_2(s))\times F^j_{ab}\left(\frac{e_1(t)+e_2(s)}{2}\right)\epsilon_{jkl}R^k_{e_1}R^l_{e_2},
\end{split}
\end{equation}
where $R^j_e$ is a right invariant vector field defined in \eqref{right and left invariant vector fields}. It is clear that this expression involves holonomies along the edges $e_1$ and $e_2$ of the graph intersecting at $v$. This justifies the choice of alignment, although zeros in the spectrum of $\hat{V}$ make its inverse not densely defined, pointing to $\{A^j_a(x),V(R_x)\}$ as a more reasonable choice to work with.

The second reason is that the uncountably infinite number of different options for setting the new edges in $\mathcal{H}_{kin}$, become countable infinitely many in $\mathcal{H}_{Phys}$ due to diffeomorphism invariance. From these \cite{Thiemann:2006cf}, except a finite number of them, the rest are all unnatural in the sense that one could also find such kind of examples\footnote{like for instance winding the new edges an arbitrary number of times around one of the edges of the graph incident at $v$ \cite{Thiemann:2006cf}} in Lattice QFT. 

\end{enumerate}

\item \textit{Ultralocality.} In \cite{nicolai2005loop} one finds an explanation of how is the action of the Hamiltonian constraint. In particular, it derives the fact that all new vertices and edges are created at $\epsilon \rightarrow 0$ distance from the pre-existing vertex, giving rise to a fractal structure. That is why the action of the Hamiltonian constraint on the spin network makes some authors call the resulting states as `dressed' \cite{Perez:2004hj}. This is in contrast with Lattice QFT, where the action of the Hamiltonian does not create new nodes, but links two already existing ones. \cite{Thiemann:2006cf} argues that although counter-intuitive, there is no reason to compare the two models (LQG and Lattice QFT) since the former is background independent and continuous, and the latter is completely discretised and background dependent.

Finally, it also argues that since two Hamiltonian operators acting on different vertices do not commute, they affect each other by modifying the place where the new vertices and edges are created. 

\item \textit{On-shell vs off-shell closure.} An important consistency requirement for the theory would be to be able to certify the closure of the quantum equivalent of the Dirac algebra \eqref{Dirac algebra} \cite{nicolai2005loop}. As we already mentioned, the main problem is that the third relation \eqref{H commutator} includes a structure function which destroys the Lie algebra character. The quantum equivalent of \eqref{D commutator} and \eqref{H D commutator} are \cite{Thiemann:2006cf}
\begin{subequations}
\begin{align}
[\hat{D}(\vec{N}),\hat{D}(\vec{N}')]=8\pi i G \hbar \hat{D}(\mathcal{L}_{\vec{N}}\vec{N}'),\label{quantum D commutator}\\
[\hat{D}(\vec{N}),\hat{H}(N')]=8\pi i G \hbar \hat{D}(\mathcal{L}_{\vec{N}}N').\label{quantum H D commutator}
\end{align}
\end{subequations}
As for the third, \eqref{H commutator}, one would seek an expression of the kind
\begin{equation}
[\hat{H}^\dagger(N),\hat{H}^\dagger(N')]=\hat{O} (N,N'),
\label{off-shell closure}
\end{equation}
where $\hat{H}^\dagger(N)$ is the dual of the Hamiltonian constraint, and $\hat{O}$ is an operator on $\mathcal{H}_{kin}$. This is what we mean here by strong closure or `off-shell closure'. In \cite{nicolai2005loop} they put a lot of emphasis in fulfilling this requirement as not only it provides a consistency check on the theory but, more importantly, it proves the quantum space-time covariance of the theory. However, at present, it is unknown how to derive explicitly the right hand side of \eqref{off-shell closure}. One must then resort to weaker notions of closure. Two are proposed: a first option is that it was shown \cite{thiemann1996anomaly,thiemann1998QSD}
\begin{equation}
\lim_{\epsilon\rightarrow 0}(\Psi\ket{\hat{O}(N,N',\epsilon) \psi}=0; \quad \hat{O}(N,N',\epsilon):=[\hat{H}(N,\epsilon),\hat{H}(N',\epsilon)],
\label{on-shell closure}
\end{equation}
for $(\Psi|\in \mathcal{D}_{Diff}^\star$, and $\ket{\psi}\in \mathcal{D}\subset \mathcal{H}_{kin}$. This is called `on-shell closure' and let us remark that the limit is taken after calculating the commutator. Since the limit and commutator do not necessarily commute, another option would be \cite{Thiemann:2006cf}
\begin{equation}
(\Psi\ket{[\hat{H}^\dagger(N),\hat{H}^\dagger(N')]\psi}=0.
\label{partly on-shell closure}
\end{equation}

This result comes from the fact that although operator $\hat{O}(N,N')$ in \eqref{off-shell closure} is unknown, its action on spin network functions is $\hat{O}(N,N')\propto (\hat{U}(\varphi)-\hat{U}(\varphi'))\hat{O}$, for $\hat{O}$ an operator on $\mathcal{H}_{kin}$. Then, making use of diffeomorphism invariance, \eqref{partly on-shell closure} holds. A different quantization of the right hand side of \eqref{H commutator} was proposed in \cite{thiemann1998QSDII,thiemann1998QSDIII,thiemann1998QSDIV,thiemann1998QSDV,thiemann1998QSDVI} that also annihilates $\mathcal{D}_{Diff}^\star$. 

That is why \cite{Thiemann:2006cf} states that for two operators $\hat{O}_1$ and $\hat{O}_2$ one will say $\hat{O}_1\sim \hat{O}_2$ provided that $\hat{O}_1-\hat{O}_2$ annihilates in $\mathcal{D}_{Diff}^\star$. 
\cite{Thiemann:2006cf} argues that although not fulfilled the off-shell closure, still we are only interested in the physical states, which in particular fulfil spatial diffeomorphism invariance. However, it is important mentioning that \cite{nicolai2005loop,nicolai2007loop} disagree and insist on the necessity of off-shell closure to ensure full spacetime covariance of the theory.

\item \textit{Semiclassical limit.}
\cite{Thiemann:2006cf} proposes that, in order to make progress with the closure of the algebra of constraints, one may try applying it to semiclassical states, by using the expectation values and substituting the quantum operators by their classical analogues (i.e. commutators by Poisson brackets and so on).

There are however two obstacles. Firstly, the volume operator appearing in the Hamiltonian is not analytically diagonalisable. Fortunately, \cite{giesel2007AQGI,giesel2007AQGII,giesel2007AQGIII} calculations with the volume  operator can be performed using coherent states, solving the first problem.
Additionally, semiclassical tools are not appropriate for operators that change the graph, such as the Hamiltonian operator, that creates new nodes and edges. Therefore, new tools are needed.

\item \textit{Solutions and inner product.} Although solutions to all constraints can be found using algorithms \cite{thiemann1998QSDII,thiemann1998QSDIII,thiemann1998QSDIV,thiemann1998QSDV,thiemann1998QSDVI}, there is no physical inner product yet.

\end{enumerate}

\section{Master constraint}

Since the problem with all the previous work regarding the Hamiltonian constraint seems to have its origin in the fact that the commutator of two Hamiltonian constraints breaks the Lie algebra character of $\mathfrak{D}$, Thiemann proposed the so called Master constraint in substitution of the Hamiltonian constraint\footnote{Another similar proposal called extended Master constraint would be adding $+q^{ab}D_aD_b$ to the numerator, where $D_a$ indicates the diffeomorphism constraint.} \cite{Thiemann:2003zv,Thiemann:2007zz}
\begin{equation}
\mathbf{M}:=\int_{\Sigma}dx^3\frac{H(x)^2}{\sqrt{\det(q)(x)}}
\label{Master constraint}
\end{equation}
It can be seen that this constraint is equivalent to the Hamiltonian constraint as
\begin{enumerate}
\item $\mathbf{M}=0 \iff H(N)=0 \quad\forall N$ 
\item $\{O,\{O,\mathbf{M}.\}\}_{\mathbf{M}=0}=0\iff \{O,H(N)\}_{H(N')=0}=0\quad \forall N,N'$.
\end{enumerate}
It is also trivial to see that $\mathbf{M}$ is spatial-diffeomorphism invariant. Then, one may substitute $\mathfrak{D}$ \eqref{Dirac algebra} for the more convenient (true Lie) algebra $\mathfrak{M}$:
\begin{subequations}
\begin{align}
&\{D(\vec{N}), D(\vec{N}')\}=8\pi G D(\mathcal{L}_{\vec{N}}\vec{N}'), \label{alternative D commutator }\\
&\{D(\vec{N}), \mathbf{M}\}=0, \label{D M commutator}\\
&\{\mathbf{M}, \mathbf{M}\}=0, \label{M commutator}
\end{align}
\end{subequations}
This algebra has several advantages:
\begin{enumerate}
\item As $\mathbf{M}$ is invariant under spatial diffeomorphisms, the corresponding quantum operator $\mathbf{\hat{M}}$ must be defined on $\mathcal{H}_{Diff}$ and not on $\mathcal{H}_{kin}$.
\item If $\mathbf{\hat{M}}$ can be defined as a self-adjoint operator, and $\mathcal{H}_{Diff}$ decomposes into the direct sum of $\mathbf{\hat{M}}$-invariant separable Hilbert spaces, $\mathcal{H}_{Phys}$ exists \cite{Thiemann:2006cf}.
\item As there is just one Master constraint, this means that it is anomaly free and we can consider the creation of new loops, for example, where there was already one. These loops that would the Hamiltonian constraint anomalous, makes the spectrum of the Hamiltonian not contain the value 0, but one can correct that substracting the minimum value of this spectrum to all eigenvalues \cite{Thiemann:2006cf}.
\item The spectrum gap (related to the normal ordering constant for free theories) is finite due to the presence of $\sqrt{\det(q)}$ in the denominator of \eqref{Master constraint} \cite{Thiemann:2006cf}.

%
%

\item Since we can now attach loops following the pre-existing graph, we get rid of the issue of ultralocality and resembles the Lattice QFT situation.
\end{enumerate}
However, as semiclassical states on $\mathcal{H}_{Diff}$ have not been constructed, the semiclassical limit of the graph changing master constraint (that creates new loops and edges) cannot be worked out. Therefore, a non graph changing master constraint is currently the only option.
\cite{Thiemann:2007zz} indicates three open questions of the canonical approach:
\begin{enumerate}
\item As the normalization factor $\eta_{[s]}$ is not indicated, there is an ambiguity from  $\braket{\cdot,\cdot}_{diff}$ that carries over to $\braket{\cdot,\cdot}_{phys}$.
\item The limit of regularization of either the Hamiltonian constraint or the master constraint is not unique.
\item We need to fix the irreducible representation we will use.
\end{enumerate}
Thiemann \cite{Thiemann:2007zz} proposes to select the correct constraint by inspecting the semiclassical regime that would allow to identify semiclassical states where classical General Relativity is realized.

\section{Why not $\gamma=\pm i$?}

We did not commented it before, but in the early times of the development of LQG, people would choose $\gamma = \pm i$ instead of the real value that it is commonly used nowadays. Using $\gamma = \pm i$, one would be using conjugate variables $E^\mathbb{C}$ and $A^\mathbb{C}$ that would become $sl(2,\mathbb{C})$ and $SL(2,\mathbb{C})$ valued respectively, and the second term of the Hamiltonian would disappear, giving as a result a Hamiltonian that if multiplied with $\sqrt{\det(q)}$, becomes $4^{th}$ order polynomial. However, to keep the theory real, one must impose reality conditions, what makes the spin connection $\Gamma$ not polynomial, and finding representation of the formal algebra $\mathfrak{V}$ unattainable (so far at least).

On the other hand, \cite{thiemann1996anomaly} explained that using a real valued $\gamma$ is the right choice as it makes the Hamiltonian have density weight one, and also showed that any background independent quantum field theory will display no UV divergences provided it has such density weight\footnote{The Hamiltonian of the standard model in flat spacetime is of density weight 2.}. This is why a real value of $\gamma$ is preferred nowadays.

\section{Coupling to matter field}

So far all we have talked about has been pure gravity sector. However, the theory should allow for couplings to Standard model too. It has been shown (see for example chapter 12 in \cite{Thiemann:2007zz} for a complete review on the topic) that the quantization poses no additional conceptual problems to the theory. The procedure consists on casting the variables in the action of Yang-Mills, fermionic and Higgs fields, into the Ashtekar language of connections and electric fluxes (as it is done with pure gravity).

To avoid similar problems as those described in the previous subsection with $\gamma=\pm i$, one will have to make use of half-densitised variables (that is, multiply the variables by $\sqrt{\det(q)}$). 
Then, we get a real connection and a Hamiltonian of weight one, what is key to avoiding divergences \cite{Thiemann:2007zz}. 
When implementing this formalism in a spin network, the description will be similar to Lattice gauge theory in that matter and fermionic fields will be attached to vertices, and gauge fields to edges \cite{nicolai2005loop}. 

However, it is not completely true that any kind of matter coupling is allowed in LQG \cite{nicolai2005loop}, as the Master operator will depend on such matter content to either have or not the zero on its spectrum \footnote{As said before, this plays a role similar to the infinite normal ordering constant in QFT}. Therefore, there might be some restriction on the matter content due to the same reasons that led to consider supersymmetry \cite{Thiemann:2007zz}.

However these and others are still open questions. In particular, although coupling to matter in spin foams is also possible \cite{Rovelli2004Quantum} (see chapter 9 in \cite{Rovelli:2014ssa} for an introduction), a derivation of the gravity-matter interaction classical limit is missing \cite{Pullin:2017}.


\section{Canonical quantization and spin foams}

There are several issues regarding the relation between the two main lines of research of Loop Quantum Gravity, and in particular whether one can derive one from the other \cite{rovelli1999projector}. Although at the beginning spin foams were proposed as evolution of spin networks in time, it has not yet been proven whether such connection exits, as the starting point for spinfoams is very different from the canonical formulation.

Other important issue pointed out in \cite{nicolai2007loop} is the fact that the Hamiltonian acts only in $1\rightarrow 3$ moves (that is, it creates 3 new nodes from an existing one), while by spacetime covariance one expects also $0\rightarrow 4$ and $2\rightarrow 2$ moves. However this is not possible, as noticed by first time in \cite{Reisenberger:1996pu} in the canonical quantization, although these moves are necessary. 

Finally, the last criticism in \cite{nicolai2007loop} is that there are many models for spinfoam gravity. Although this might have been true some time ago, there is general consensus on the loop community that there is a model right now, called EPLR and the one that we described here, that has the chance of being the right model. In fact it provides the correct classical limit of GR \cite{Rovelli:2014ssa} (at least under some circumstances).

\section{Other problems}

Finally there are other problems that need to be addressed too. \cite{nicolai2005loop} mentions some
\begin{enumerate}
\item It is desirable that the theory is able to explain the 2-loop order divergence, and why the counterterm necessary for obtaining a finite S matrix must be \cite{goroff1985quantum,goroff1986ultraviolet,van1992two}
\begin{equation}
\Gamma^{(2)}_{div}=\frac{1}{\epsilon}\frac{209}{2880}\frac{1}{(16\pi^2)^2}\int d^4x\sqrt{g}C_{\mu\nu\rho\sigma}C^{\rho\sigma\lambda\tau}C_{\lambda \tau}^{\mu\nu}
\end{equation}
\item It might be possible to add higher order terms to the Hamiltonian such that it still fulfils the consistency requirements.
\item Although we have seen that kinematical geometrical operators have a discrete spectrum, it is not clear if this is the case for dynamical operators (at least in the canonical formulation).
\item In quantum mechanics the notion of `unitarity' is very important. However, since in LQG there is no time a priory, it is difficult to say whether it will still be a true aspect of Physics \cite{nicolai2005loop}.
\end{enumerate}

\section{Conclusions}
In this chapter we have seen that there are still some problems regarding the formulation of loop quantum gravity, that need to be addressed. The main focus was on the Hamiltonian constraint and the problem with space time covariance, that is whether the algebra of constraints closes or not off-shell. We have seen that most of these problems can be overcome using a master constraint, although this does not completely solve all the issues. Finally, we have also reviewed some of the points concerning matter coupling, connection to spin foams and other open problems that remain open.
\chapter{Conclusions}\label{ch:conclusions}
In this work we have very briefly reviewed the field of Loop Quantum Gravity. Unfortunately, although this is just the basics of the theory, we do not have pages to go further (sorry). Beyond this dissertation there is a lot of work to understand the theory better. From the more mathematical and conceptual lines all the way down to applications, there is a broad range of lines of research. For example, the sum over different spin foams seems to be a natural extension of the covariant formalism. In this line, Group Field Theory techniques are currently under research. 

Other important focus of research is, as we briefly mentioned, the search for the continuum and the classical limits. Here a number of different techniques are being applied, such as RG flow, or coherent states. It is expected that these developments will allow for consistency checks of the theory.

However, no new physical theory will be accepted until it is able to make predictions, and they prove right. That is why the field of applications is one of the most active ones. Obviously, these mainly lie in the study of the behaviour of black holes, and the study of Loop Quantum Cosmology. 
In the black hole section, there has been important progress and the entropy of different kinds of black holes has been calculated. There is also new work exploring holography. It has even been proposed the existence of white holes as the result of a bounce back of black holes after the collapse, forming a Planck star in between \cite{rovelli2014planck}. This is interesting since the wavelength would be in the order of $10^{-14}$ cm, which is well in the detection range of our instruments, and could be related to the poorly understood fast Gamma Ray Bursts.

The other main field of research has been Loop Quantum Cosmology. Here maybe one of the most significant predictions is that big bang was in reality a big bounce. This research field is very active and is currently exploring different models with various conditions of homogeneity and isotropy. 

There is finally some work being carried out to try to extract data from observations to contrast predictions. For instance, the Cosmic Microwave Background can be used to fix the value of the Immirzi parameter within certain error, and this value can be checked against the one needed to obtain the Hawking entropy formula right. In fact, within the current error bars, the results for $\gamma$ are consistent \cite{Pullin:2017}. 

The information on these topics is abundant. For example, for an introduction to LQG at a masters level I would recommend both \cite{Rovelli:2014ssa} for the covariant formalism and \cite{Perez:2004hj} for the canonical basics. Also, for a state of the art and latest developments information, the book \cite{Pullin:2017} constitutes an excellent source. Therefore, the main structure of this dissertation was obtained from these references, and for the open issues chapter, from \cite{nicolai2005loop,nicolai2007loop,Thiemann:2006cf}. Finally, most of the wisdom from the canonical approach can be found in \cite{Thiemann:2007zz}, although it is highly technical.

So, it is my personal belief that this is an interesting theory to be explored. Clearly, different approaches to quantum gravity such as string theory are also very necessary. I think the key is just understanding what are the weak points of each theory. But as said in \cite{nicolai2005loop}, since we still do not have a single fully finished theory of quantum gravity, all different approaches are worth attention. It may well be that everyone is wrong, but if we do not try, how will we know? So let us just be curious.


\appendix
\include{appendix1}
\include{appendix2}
\begin{romanpages}

\addcontentsline{toc}{chapter}{Bibliography}

\bibliography{bibliography}        
\bibliographystyle{unsrt}  
\end{romanpages}
\end{document}